\documentclass[fleqn,usenatbib]{mnras}

\usepackage[T1]{fontenc}
\usepackage{ae,aecompl}
\usepackage[utf8]{inputenc}
\usepackage[english]{babel}
\usepackage{rotating}
\usepackage{multirow}
\usepackage{url}
\usepackage{diagbox}
\usepackage{tikz,wrapfig}

\usepackage{latexsym}	
\usepackage{array}
\newcolumntype{P}[1]{>{\centering\arraybackslash}p{#1}}

\usepackage{mathtools}
\usepackage{xcolor}

\usepackage{amsmath,bbm,bm,amssymb,mathtools,tikz}
\usepackage{soul}
\usepackage{gentium}
\usepackage{graphicx,psfrag,epsf,natbib,subfig,wrapfig}
\usepackage{booktabs}
\usepackage{multirow}
\usepackage[export]{adjustbox}

\usepackage{verbatim}
\usepackage{booktabs}
\usepackage{xcolor}
\usepackage{tikz}

\usepackage{algorithmic,algorithm}
\usepackage{hyperref}
\usepackage{url}

\newcommand*{\boldone}{\text{\usefont{U}{bbold}{m}{n}1}}

\newcommand{\bLambda}{\boldsymbol{\Lambda}}
\newcommand{\bDelta}{\boldsymbol{\Delta}}
\newcommand{\bPhi}{\boldsymbol{\Phi}}

\newcommand{\bSigma}{\boldsymbol{\Sigma}}
\newcommand{\bPsi}{\boldsymbol{\Psi}}

\newcommand{\bTheta}{\boldsymbol{\Theta}}
\newcommand{\bOmega}{\boldsymbol{\Omega}}

\newcommand{\bD}{\boldsymbol{D}}
\newcommand{\bff}{\boldsymbol{f}}
\newcommand{\bF}{\boldsymbol{F}}
\newcommand{\bV}{\boldsymbol{V}}
\newcommand{\bg}{\boldsymbol{g}}

\newcommand{\bI}{\boldsymbol{I}}

\newcommand{\bk}{\boldsymbol{k}}

\newcommand{\bT}{\boldsymbol{T}}

\newcommand{\be}{\boldsymbol{e}}

\newcommand{\bq}{\boldsymbol{q}}
\newcommand{\bQ}{\boldsymbol{Q}}
\newcommand{\bR}{\boldsymbol{R}}
\newcommand{\bW}{\boldsymbol{W}}
\newcommand{\bu}{\boldsymbol{u}}
\newcommand{\bv}{\boldsymbol{v}}

\newcommand{\bx}{\boldsymbol{x}}
\newcommand{\bX}{\boldsymbol{X}}

\newcommand{\bY}{\boldsymbol{Y}}
\newcommand{\bZ}{\boldsymbol{Z}}

\newcommand{\bzero}{\boldsymbol{0}}

\newcommand{\mN}{\mathcal N}

\newcommand{\br}{\boldsymbol{r}}

\renewcommand{\Pr}{{\mathbb P}\textnormal{r}}

\let\muold\mu
\renewcommand{\mu}{{\bm\muold}}
\let\LambdaOLD\Lambda
\renewcommand{\Lambda}{{\bm\LambdaOLD}}
\let\Psiold\Psi
\renewcommand{\Psi}{{\bm\Psiold}}
\let\Usiold\Upsilon
\renewcommand{\Upsilon}{{\bm\Usiold}}
\let\Sigmaold\Sigma
\renewcommand{\Sigma}{{\bm\Sigmaold}}
\let\Gammaold\Gamma
\renewcommand{\Gamma}{{\bm\Gammaold}}
\let\sigmaold\sigma
\renewcommand{\sigma}{{\bm\sigmaold}}

\newcommand{\Tr}{\mathrm{Tr}~}
\def\shalf{\mbox{{\footnotesize$\frac{1}{2}$}}}
\newcommand{\normal}{{\mathcal N}}

\newcommand{\bbR}{\mathbb{R}}

\newtheorem{result}{Result}
\newtheorem{remark}{Remark}

\usepackage[many]{tcolorbox}
\newtcolorbox{mybox}[1][]{
    tikznode boxed title,
    enhanced,
    arc=0mm,
    interior style={white},
    attach boxed title to top center= {yshift=-\tcboxedtitleheight/2},
    fonttitle=\bfseries,
    colbacktitle=white,coltitle=black,
    boxed title style={size=normal,colframe=white,boxrule=0pt},
    #1}

\definecolor{err1}{RGB}{27,158,119}
\definecolor{err2}{RGB}{217,95,2}
\definecolor{err3}{RGB}{117,112,179}
\definecolor{err4}{RGB}{231,41,138}
\definecolor{err5}{RGB}{166,118,29}
\definecolor{err6}{RGB}{102,102,102}

\definecolor{sim1}{RGB}{117,   112,  179}
\definecolor{sim2}{RGB}{231,    41,  138}
\definecolor{sim3}{RGB}{102,   166,   30}
\definecolor{sim4}{RGB}{102,   194,  165}
\definecolor{sim5}{RGB}{252,   141,   98}

\definecolor{comp1}{RGB}{166,   118,   29}
\definecolor{comp2}{RGB}{102,   102,  102}

\title[Latent characterisation of gamma ray bursts]{Latent characterisation of the complete BATSE gamma ray bursts catalogue using Gaussian mixture of factor analysers and model-estimated overlap-based syncytial clustering}

\author[Dai and Maitra]{
  Fan Dai,$^{1}$\thanks{E-mail: fand@mtu.edu (FD)}
  and Ranjan Maitra,$^{2}$\thanks{E-mail: maitra@iastate.edu (RM)}\\
$^{1}$ Department of Mathematical Sciences, Michigan Technological University, 1400 Townsend Dr, Houghton, MI 49931, USA\\
$^{2}$ Department of Statistics, Iowa State University, 2438, Osborn Drive, Ames, Iowa 50011-1090, USA
}
\date{Accepted 2024 November 11. Received 2024 November 11; in original form 2024 September 15}

\pubyear{2023}

\begin{document}
\label{firstpage}
\pagerange{\pageref{firstpage}--\pageref{lastpage}}
\maketitle

\begin{abstract}
	Characterising and distinguishing gamma-ray bursts (GRBs) has interested astronomers for many decades. While some authors have found two or three groups of GRBs by analyzing only a few parameters, recent work identified five ellipsoidally-shaped groups upon considering nine parameters $T_{50}, T_{90}, F_1, F_2, F_3, F_4, P_{64}, P_{256}, P_{1024}$. Yet others suggest sub-classes within the two or three groups found earlier. Using a mixture model of Gaussian factor analysers, we analysed 1150 GRBs, that had nine parameters observed, from the current Burst and Transient Source Experiment (BATSE) catalogue, and again established five ellipsoidal-shaped groups to describe the GRBs. These five groups are characterised in terms of their average duration, fluence and spectrum as shorter-faint-hard, long-intermediate-soft, long-intermediate-intermediate, long-bright-intermediate and short-faint-hard. The use of factor analysers in describing individual group densities allows for a more thorough  group-wise characterisation of the parameters in terms of a few latent features. However, given the discrepancy with many other existing studies that advocated for two or three groups, we also performed model-estimated overlap-based syncytial clustering (MOBSynC) that successively merges poorer-separated groups. The five ellipsoidal groups merge into three and then into two groups, one with GRBs of low durations and the other having longer duration GRBs. These groups are also characterised in terms of a few latent factors made up of the nine parameters. Our analysis provides context for all three sets of results, and in doing so, details a multi-layered characterisation of the BATSE GRBs, while also explaining the structure in their variability.

\end{abstract}




\section{Introduction}
\label{sec:intro}
The astrophysics community has long been interested in the phenomenon of gamma ray bursts (GRBs) that are the most energetic and intense electromagnetic radiations originating from massive stars in the space, because they are believed to provide insights into star formation and the birth of the early universe \citep{piran92,piran05,iokaetal16}.
In general, investigations about GRBs  have focused mainly on unravelling their origins by characterising the properties and features of these observed cosmic explosions, with many fundamental results suggesting the existence of sub-populations inside the GRBs \citep{mazetsetal81,norrisetal84,dezalayetal92,yangetal22,steinhardtetal23,zhuetal24}. \cite{kouveliotouetal93} analysed 222 GRBs observed from the Burst and Transient Source Explorer (BATSE) 1B catalogue and identified bimodality in the distribution of the duration variable (specifically, $T_{90}$ in the $\log_{10}$ scale). The parameter $T_{90}$, or the time by which 90 per cent of the flux arrives, is the most prominent single attribute of a GRB, and was used to classify  GRBs into two kinds: those with a short ($T_{90} < 2s$) and long duration ($T_{90}\geq 2s$). This classification indicated distinct astrophysical origins of GRBs with the short duration bursts believed to be associated with the merger of neutron stars, either with another neutron star, or with a black hole~\citep{nakar07, berger2013,Ghirlanda2017}, while the longer duration bursts were surmised to be the progenitors of the collapse of massive stars \citep{paczynski98,woosleyandbloom06,stanek2008,pendletonetal97}. Subsequent investigations~\citep{horvathetal98} pointed towards the existence of 
an intermediate-duration category for the BATSE 3B data, which was further supported by studies that proposed three Gaussian components for the distribution of durations \citep{horvath02,hakkilaetal03,horvathetal04,horvath09,hujaetal09,tarnopolski15,zitounietal15,horvathandtoth16}. The existence of a third GRB group was also supported by multivariate analysis and non-parametric hierarchical and $k$-means clustering  performed by \citet{mukherjeeetal98} and \citet{chattopadhyayetal07}. 
Their investigations used three observed and three composite parameters, with the observed parameters being $T_{50}$ or the time by which 50 per cent of the flux arrives, $T_{90}$, $P_{256}$ or the peak flux in a bin of 256 ms, and the three composite parameters as the total fluence or $F_t$ that is the sum of the four time-integrated fluences ($F_1,F_2,F_3,F_4$) in the  20-50, 50-100, 100-300, and $>$ 300 keV spectral channels, $H_{32}= F_3/F_2$ or spectral hardness ratio using $F_2$ and $F_3$, and $H_{321}= F_3/(F_1 + F_2)$, or spectral hardness based on the ratio of channel fluences $F_1,F_2,F_3$. 
In subsequent work that also stressed the importance of proper initialisation and careful application of clustering algorithms in astrophysics, \citet{chattopadhyayandmaitra17} demonstrated that analysis of the six parameters used in \citet{mukherjeeetal98} and \citet{chattopadhyayetal07} shows that  the optimal Gaussian Mixture Model (GMM) model-based clustering (MBC) solution has five ellipsoidally-dispersed GRB groups in the BATSE 4Br catalogue. Further detailed work~\citep{chattopadhyayandmaitra18} illustrated that all nine original parameters are directly involved in clustering, and provided further evidence of five ellipsoidally-shaped groups by performing MBC with multivariate-$t$ distributions that allowed for the characterisation of heavier-tailed distributions. \citep[However, ][suggested these groups are only cut groups of the previously well-known three types of GRB groups]{tothetal19}.  Subsequent additional analyses~\citep{berryandmaitra19,almodovarandmaitra20} have confirmed these findings of five distinct groups of GRBs in the BATSE 4Br catalogue. In particular, \citet{berryandmaitra19} demonstrated how restricting to only a couple of parameters (indeed, only the duration $T_{90}$) results in a two- or three-groups solution. Indeed, and unsurprisingly, further work~\citep{tarnopolski22,bhaveetal22,tarnopolski19} using only the duration and hardness parameters in the analysis has continued to identify only two groups, with a somewhat unclear possibility for a third group of GRBs in the BATSE dataset.

The complete BATSE GRB catalogue, publicly available from \url{https://heasarc.gsfc.nasa.gov}, comprises the gamma-ray bursts detected by the BATSE instrument on the Compton Gamma-Ray Observatory (CGRO). It includes the 1635 gamma-ray bursts from the BATSE 4B Catalog (triggers 105 through 5586, observed between April 19, 1991, and August 29, 1996) as well as 1067 triggered bursts since the publication of the BATSE 4B Catalog. There are 2702 such records, out of which only 1971 records are on the nine parameters of interest in this article. (The GRBs with trigger numbers 
2463 and 2464 that were included in the BATSE 4Br catalogue are no longer included in the current complete catalogue.) As in earlier versions of the catalogue, there are recorded zeroes for some of these parameters, and private communication from Charles A. Meegan of the BATSE GRB team in ~\citet{chattopadhyayandmaitra17} indicated that these zeroes are not numerical, but rather should be considered as missing values. Therefore, 373 of these 1971 GRBs are incomplete (mostly in the fluence parameters), and excluding them provides in 1598 complete GRBs with all nine parameters observed. 

\begin{figure}
\centering
    \includegraphics[width = \columnwidth]{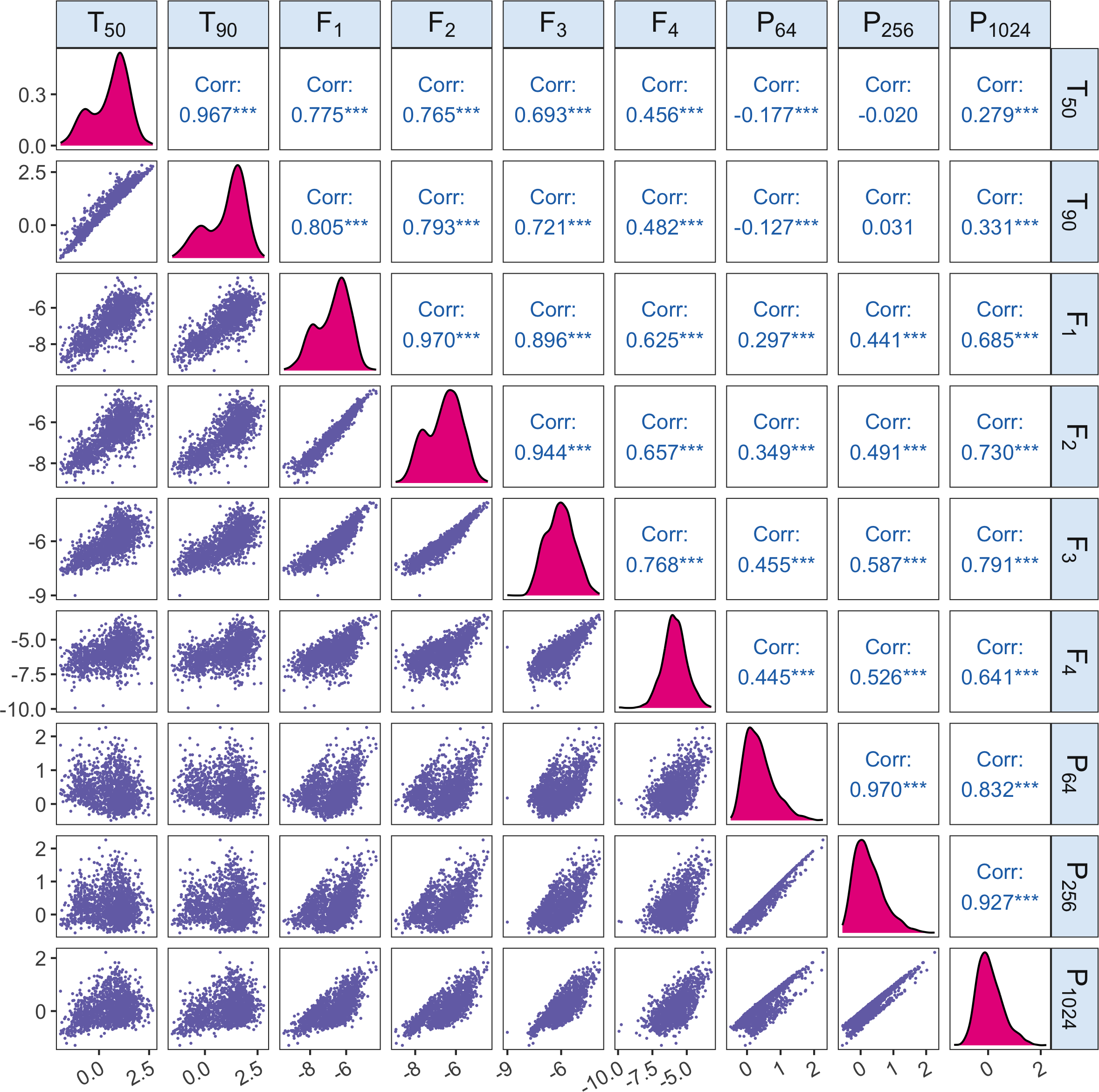}
    \caption{Densities and scatter plots of the nine features (after $\log_{10}$ transformation) for the 1598 complete BATSE Catalog GRBs. Correlations between features are shown in the upper panel.}
    \label{fig:grb}
\end{figure}

\begin{figure}
\centering
\mbox{\subfloat[Flux arrival time\label{fig:grb-se-t}]{\includegraphics[width=0.3\columnwidth]{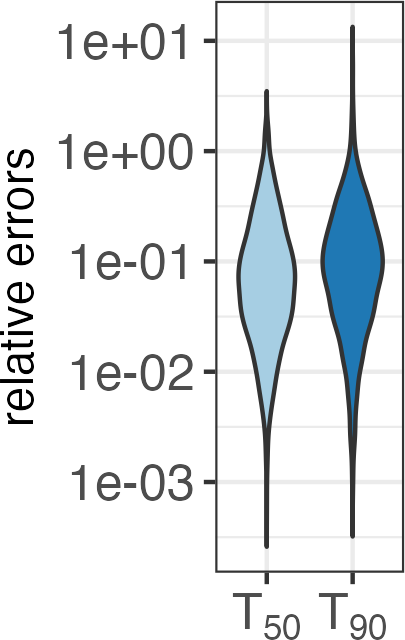}
    }
    \subfloat[Time-integrated fluence\label{fig:grb-se-f}]{\includegraphics[width=0.37\columnwidth]{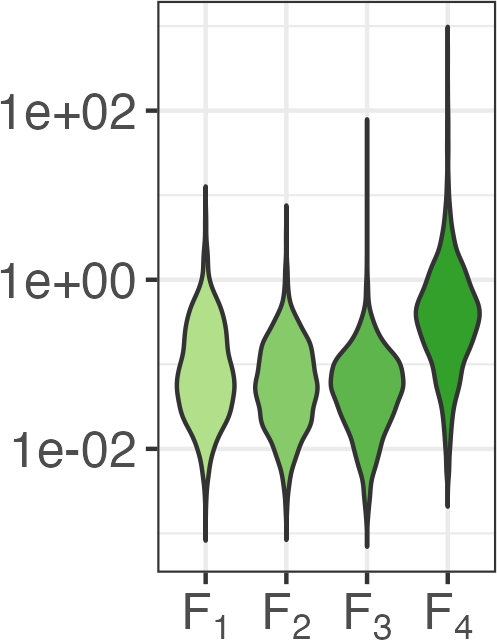}
    }
    \subfloat[Peak flux\label{fig:grb-se-p}]{\includegraphics[width=0.325\columnwidth]{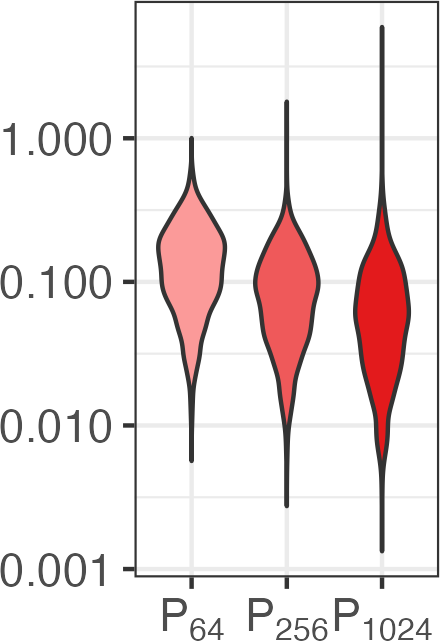}
    }
}
\caption{Violin plots of the relative errors (in $\log_{10}$ scale) of all nine parameters: (a) two flux arrival times (b) four time-integrated fluences, and (c) three peak fluxes, for the 1598 complete BATSE Catalog GRBs.}
\label{fig:me}
\end{figure}

An aspect of note is that the nine original parameters, that is, the $T_{50}$, $T_{90}$, $F_1$, $F_2$, $F_3$, $F_4$ mentioned above and $P_{64}$, $P_{256}$ and $P_{1024}$ or the three peak fluxes in time bins of $64, 256$ and $1024$ milliseconds are heavily right-skewed while their $\log_{10}$-transformed versions, with the transformation done to alleviate their individual skewness, exhibit a high degree of correlation (Figure \ref{fig:grb}). Additionally, these 1598 GRBs are observed with individually estimated measurement errors that are heavily skewed, and range in the interval $0.001{-}27.456s$ for $T_{50}$, $0.001{-}61.576s$ for $T_{90}$, $7.439\times10^{-10}{-}3.347\times10^{-7}\mbox{ erg}/cm^2$ for $F_1$, $8.808\times10^{-10}{-}2.932\times10^{-7}\mbox{ erg}/cm^2$ for $F_2$, $2.74\times10^{-9}{-}1.159\times10^{-6}\mbox{ erg}/cm^2$ for $F_3$, $2.05\times10^{-8}{-}5.474\times10^{-6}\mbox{ erg}/cm^2$ for $F_4$, $0.158{-}2.63\mbox{ photons}/cm^2$ for $P_{64}$, $0.076{-}1.292 \mbox{ photons}/cm^2$ for $P_{256}$ and $0.035{-}1.29\mbox{ photons}/cm^2$ for $P_{1024}$. Figure~\ref{fig:me} displays the distributions of the relative errors (in $\log_{10}$ scale) and Table~\ref{tab:me} lists the numbers of GRBs whose observed values are greater than $1\sigmaold$, $2\sigmaold$ and $3\sigmaold$ regarding each of the nine parameters. To keep non-negligible observations compared to their measurement errors and meanwhile, ensure a sufficient sample size for clustering, we determine $2\sigmaold$ as the acceptance level for the two flux arrival times, and $1\sigmaold$ for the four time-integrated fluences and the three peak fluxes, which reduces the GRB dataset from 1598 records to 1150 bursts that form our thinned sample that is used in our analysis. (These measurement errors provide important information on the observed parameters, and should ideally be integrated in any analysis, but we do not yet have validated statistical methodology to do so, and so we only consider the thinned sample in our analysis.)
\begin{table*}
\centering
\caption{Numbers of GRBs with observed values above the acceptance levels according to each of the nine parameters.}
\label{tab:me} 
\begin{tabular}{c|c|c|c|c|c|c|c|c|c}
\hline\hline
\diagbox[width=16em]{\bf{Acceptance level}}{\bf{Feature}} 
&\bm{$T_{50}$}& \bm{$T_{90}$} & \bm{$F_1$}& \bm{$F_2$}& \bm{$F_3$}& \bm{$F_4$}& \bm{$P_{64}$}& \bm{$P_{256}$}& \bm{$P_{1024}$}\\
\hline\hline
\bm{$1\sigma$} 
& $1577$ &$1571$      &$1555$      &$1584$      &$1593$      &$1276$    &$1598$     &$1597$      &$1597$
\\
\hline
\bm{$2\sigma$} 
& $1519$ &$1501$      &$1467$      &$1563$      &$1581$       &$992$    &$1587$    &$1597$      &$1594$
\\
\hline
\bm{$3\sigma$}  
& $1425$ &$1387$      &$1351$      &$1512$      &$1558$       &$777$    &$1521$     &$1587$      &$1583$
\\
\hline
\end{tabular}
\end{table*}
Turning our attention back to Figure~\ref{fig:grb}, we note that while the high correlations between the nine parameters in the $\log_{10}$ scale may make it seem appropriate (and convenient) to drop most of them in our analysis, \citet{chattopadhyayandmaitra18} established that all nine parameters contain clustering information and should be included in a MBC solution. However, it is also possible that there may be a few latent variables that are linear combinations of these parameters and that can better characterise the variability in these parameters. These latent variables are called factors, and we investigate the use of these latent variables in describing our  model. 

The idea of using factor analysis in describing GRBs is not new. Indeed,  \citet{bagolyetal09} applied factor analysis on 197 long duration BATSE GRBs and demonstrated that the nine parameters in these bursts are explainable by three latent variables. In our analysis here, we incorporate the factor structure in describing each cluster in the mixture-based model used in clustering. Our findings here also establish five ellipsoidally-shaped groups, with the nine parameters in each group expressed differentially by five factors, or five latent variables.
A second aspect that we study here is that almost all the clustering algorithms applied to GRB data have assumed and thus provided ellipsoidally-shaped, that is, ellipsoidally-dispersed clusters. For instance, it is possible that the five clusters found in \citet{chattopadhyayandmaitra17,chattopadhyayandmaitra18,berryandmaitra19,almodovarandmaitra20} or here, are really sub-classes of a smaller number of general-shaped clusters as has been surmised in the literature \citep{bhardwajetal23}. We study this possibility by adapting the syncytial clustering framework of \citet{almodovarandmaitra20} to our GMM-based framework to allow for model-based overlap-based clustering (MOBSynC). Indeed, we see that the BATSE GRBs can really be characterised in a multi-layered fashion by two clusters at a coarse level, that can be further subdivided into (two for the first cluster, and three for the second cluster) ellipsoidally-shaped subgroups at the finer level. This type of multi-layered characterisation was introduced into astrophysical applications by~\citet{chattopadhyayetal22}  to characterise hot stellar systems but has not previously been applied to GRBs, and indeed never in conjunction of mixtures of factor analysers. 


The remainder of this paper is organised as follows. Section \ref{sec:meth} provides a detailed illustration on constructing the Gaussian mixture of factor analysers (Sections \ref{sec:gmm} and \ref{sec:fa}), implementing the profile likelihood maximisation to estimate the covariance parameters and determining merging of clusters using overlap measures (Section \ref{sec:merge}). Section \ref{sec:app} applies our methodology on the fully-observed GRBs from the thinned complete BATSE catalogue to detect optimal groups to classify GRBs and find latent factors to demonstrate the original features of GRBs. Section \ref{sec:con} summaries the article and discusses further work. 
Our paper also has an appendix that details the steps in fast parameter estimation in the Gaussian Mixture of Factor Analysers model.

\renewcommand{\hat}{\widehat}
\section{Statistical Framework for Clustering}
\label{sec:meth}

In this section, we detail the statistical background for the clustering algorithms used in this paper. The approach starts with MBC of data using a Gaussian mixture of factor analysers, where we also develop fast computational methods for parameter estimation. Based on the initial clustering results, we adapt the procedure described in \citet{chattopadhyayetal22} that further merges identified clusters according to generalised and pairwise overlap measures. Our methods are implemented using our {\tt MixFAD} (Mixture of Factor Analysers in Data) program and the {\tt MOBSynC} function (available at \url{https://github.com/fanstats/MixFAD-GRB}), both coded in the open-source statistical software {\tt R}~\citep{R} and its relevant packages.

\subsection{Clustering with Gaussian Mixture of Factor Analysers}
\subsubsection{Gaussian Mixture Model-based Clustering}
\label{sec:gmm}

The MBC methodology~\citep{anderson03,mclachlanandpeel00,mardiaetal06,melnykovandmaitra10} provides a principled and formal approach to grouping observations in an unsupervised setting, with data assumed to arise from a finite mixture model, that is, a mixture of component distributions, such as the multivariate Gaussian, in which case we have a GMM \citep{chattopadhyayandmaitra17}. 
Formally, our modeling and estimation framework assumes that we have $p$-dimensional observations $\bx_i,i=1,2,\dots,n$ from a GMM with $K$ components, where $\bx_i$ is generated from the $k$th component with probability $\eta_k, k=1,2,\dots,K$. Then, the loglikelihood function is \citep{day69}
\begin{equation}\label{eqn:gmm-loglik}
\ell(\bTheta;\bX) = \sum_{i=1}^{n}\log\Big\{\sum_{k=1}^{K}\eta_{k}\bphi_(\bx_i;\bmu_k,\bSigma_k)\Big\},
\end{equation}
with the data matrix $\bX = [\bx_1^\top,\bx_2^\top,\dots,\bx_n^\top]^\top$, the parameter space $\bTheta = (\eta_1,\eta_2,\dots,\eta_K,\bmu_1,\bmu_2,\dots,\bmu_K,\bSigma_1,\bSigma_2,\dots,\bSigma_K)$, and the Gaussian probability density function (PDF) $\bphi(\cdot)$ for the $k$th mixture component.
Parameter estimation by optimising \eqref{eqn:gmm-loglik} is almost always intractable, but made possible by the expectation-maximisation (EM) algorithm \citep{dempsteretal77, rubinandthayer82, mclachlanandkrishnan08}, and specifically, by the  introduction of a latent group indicator variable $z_i$ for each $\bx_i$, such that the conditional distribution of $\bx_i$ given $z_i=k$ is $\normal_k(\bmu_k,\bSigma_k)$ and $\mathrm{{\mathbb P}r}(z_i=k) = \eta_k$. Then, for $(\bX,\bZ)$ where $\bZ = (z_1,z_2,\dots,z_n)$, and with $\boldone(\cdot)$ denoting the indicator function, we can obtain \citep{mclachlanandpeel00} the loglikelihood function for the complete (observed as well as augmented) data as 
\begin{equation}
\begin{split}
\label{eqn:gmm-emloglik}
\ell(\bTheta;\bX,\bZ) = &\sum_{i=1}^{n}\sum_{k=1}^{K}\boldone(z_i=k)\Big\{\log\eta_k+
\log \phi(\bx_i;\bmu_k,\bSigma_k)\Big\}.
\end{split}
\end{equation}
The EM algorithm starts with some initial values of $\bTheta$, and then iterates between the E- or the expectation step  and M- or maximisation step until the updates converge to a solution that is shown to be a maximum likelihood estimate. Specifically, we have the E- and M-steps as follows:
\paragraph*{E-step:} For the unknown term $\boldone(z_i=k)$ in \eqref{eqn:gmm-emloglik}, compute its conditional expectation $\mathbb{E}{[\boldone(z_i=k)|\mathbf{X};\bTheta]}$, and obtain the expected complete loglikelihood (the so-called $Q$-function) as
\begin{equation}\label{eqn:gmm-qfun}
\begin{split}
    \mathrm{Q}(\bTheta^*;\mathbf{X},\bTheta) 
         = &\sum_{i=1}^{n}\sum_{k=1}^{K}\mathrm{E}{[\boldone(z_i=k)|\mathbf{X};\bTheta]}\Big\{\log\eta_k+     \\
	   &\qquad\qquad\qquad\qquad\qquad\qquad\log \phi(\bx_i;\bmu_k,\bSigma_k)\Big\},
\end{split}
\end{equation}
where 
\begin{equation}
{\mathbb E}{[\boldone(z_i=k)|\mathbf{X};\bTheta]} = \displaystyle{\frac{\eta_{k}\bphi(\bx_i;\bmu_{k},\bSigma_{k})}{\sum_{k=1}^{K}\eta_{k}\bphi(\bx_i;\bmu_{k},\bSigma_{k})}}.
\label{eqn:map}
\end{equation}
\paragraph*{M-step:} Compute the updated estimates of $\bTheta^*$ by maximising the $Q$-function \eqref{eqn:gmm-qfun}. Specifically, we obtain
\begin{equation}
\label{eq:est}
    \begin{split}
     \hat\eta_{k}^* =&
        \displaystyle{\frac{n_{k}}{n}},\\
     \hat\bmu_{k}^* =& \displaystyle{\frac{\sum_{i=1}^{n}\mathrm{{\mathbb P}r}(z_i=k|\bx_i;\bTheta)\bx_i}{n_{k}}},\\
     \hat\bSigma_{k}^* =& \displaystyle{\frac{\sum_{i=1}^{n}\mathrm{{\mathbb P}r}(z_i=k|\bx_i;\bTheta)(\bx_i-\bmu_{k}^*)(\bx_i-\bmu_{k}^*)^\top}{n_{k}}},
    \end{split}
\end{equation}
where $n_{k} = \sum_{i=1}^{n}\mathbb{E}{[\boldone(z_i=k)|\mathbf{X};\bTheta]}$.

As discussed and displayed in \citet{maitra09} and \citet{chattopadhyayandmaitra17}, initialisation is an important contributor to the optimisation performance of the EM algorithm since the convergence is to a (local) maximum in the vicinity of the initialiser. We adapt the stochastic initialisation methods of \citet{maitra13} or \citet{gorenandmaitra22} by initialising the EM algorithm from $m_1$ randomly chosen initial values and running it for a while, and then choosing, from among them, $m_2$ candidates with the highest observed loglikelihood functions after a few iterations, that are then used to finally initialise the EM and to run it to convergence. The optimal initialisation is the one that yields the largest final observed  loglikelihood value from among the $m_2$ runs. In our analysis of the GRBs in Section~\ref{sec:app}, we used $m_1=1000, m_2=10$.

Having obtained an EM-estimated maximum likelihood estimator (MLE) of $\Theta$, we classify each observation $\bx_i$ into the $k$ for which \eqref{eqn:map} is maximum. With our discussion of GMM and MBC in place, we now extend the above setup to the Gaussian Mixture of Factor Analysers framework.

\subsubsection{Gaussian Mixture of Factor Analysers MBC}
\label{sec:fa}
Factor models allow for the variability in a $p$-dimensional dataset to be explained by a few ($q<p$) underlying factors. 
In general, a factor model~\citep{thurstone31,thurstone35,anderson03} for $\bx_i$ can be represented by the linear equation
\begin{equation}
    \bx_i = \bmu + \bLambda\bF_i +\bepsilon_i,
\end{equation}
where  $\bF_i\sim\normal(\boldsymbol{0},\bI_q)$ denotes the $q$ latent factors, and $\bepsilon_i\sim\normal(\boldsymbol{0},\bPsi)$, with the so-named {\em matrix of uniquenesses} $\bPsi$ that is diagonal and stores the variances unique to the individual variables. Further, 
$\bLambda$ is a $p\times q$ factor loading matrix that explains for the $k$th feature in $\bx_i$, the weight or proportion that is contributed to it by the $j$th latent variable or factor, where $k=1,2,\ldots, p$ and $j=1,2,\ldots,q$.  For identifiability of the model, and also by nonsingularity of $\bSigma$, we constrain $q<\min(n,p)$ and $(p-q)^2>p+q$. By  construction, $\bx_i\sim\normal(\bmu,\bSigma)$ where $\bSigma = \bLambda\bLambda^\top + \bPsi$. As mentioned earlier, \citet{bagolyetal09} introduced factor analysis in astrophysics to characterise 197 long duration GRBs in the BATSE catalogue. Unlike principal components~\citep{bagolyetal98} which provides for a lower-dimensional representation of the data and that therefore assumes a singular $\bSigma$ (of rank equal to the number of principal components), a factor model has a nonsingular $\bSigma$ that can be represented by a fewer number ($pq+q$)  of parameters than the unconstrained $\bSigma$ that has $p(p+1)/2$ parameters.

In the clustering context, we extend the factor model to represent each group. Specifically, for the GMM defined in Section~\ref{sec:gmm} we generalise the definition of a factor model by specifying each Gaussian component $\phi(\bx;\bmu_k,\bSigma_k)$ in terms of a common-$q$ factor model. Thus, we get a Gaussian mixture of factor analysers \citep{mclachlanandpeel00} where the $k$th group covariance matrix has a constrained structure as $\bSigma_k = \bLambda_k\bLambda_k^{\top} + \bPsi_k$, with $\bLambda_k$ as the $p\times q$ factor loading matrix with a common rank $q$ for all components.  Consequently, the group-wise factor model is
\begin{equation}
	\label{eq:gmmfad}
	\bx_i = \bmu_k+\bLambda_k\bF_{ik} + \bepsilon_{ik},\quad \mbox{when } z_i = k,
\end{equation}
with ${\bF}_{ik}\sim\normal(\boldsymbol{0},\bI_q)$ and ${\bepsilon}_{ik}\sim\normal(\boldsymbol{0},\bPsi_k)$. Thus, in the case of GMM with factor model components, our setup is as in \eqref{eqn:gmm-loglik} with the additional refinement that the $\bSigma_k$ in each component is decomposed as $\bSigma_k=\bLambda_k\bLambda_k^\top + \bPsi_k$, yielding the observed loglikelihood function
\begin{equation}\label{eqn:gmmfad-loglik}
\ell(\bTheta;\bX) = \sum_{i=1}^{n}\log\Big\{\sum_{k=1}^{K}\eta_{k}\phi_(\bx_i;\bmu_k,\bLambda_k\bLambda_k^\top+\bPsi_k)\Big\},
\end{equation}
The representation~\eqref{eq:gmmfad} has an identifiability issue in estimation: notably, for any orthogonal matrix $\mathbf{T}_k$, the factor loading matrix $\bLambda_k$ and its orthogonal transformation $\bLambda_k \mathbf{T}_k$ yield the same likelihood value, so a constrain on the parameters is  required to ensure an unique solution of $\bLambda_k$. It is common to adopt, as we do here, a scale-invariant constraint such that the so-called {\it signal matrix} $\bLambda_k^\top\bPsi_k^{-1}\bLambda_k$ is diagonal with decreasing diagonal entries. This constraint can have $\bLambda_k$ profiled out for a given $\bPsi_k$, providing computational convenience \citep{lawley40,daietal20}.

Maximum likelihood parameter estimation can, as in Section~\ref{sec:gmm}, be achieved by employing the EM algorithm. The E-step is pretty much unchanged from that outlined there, while the M-step is, barring minor modifications, also largely the same for $\eta_k$'s and $\bmu_k$. Efficient parameter estimation of $\bLambda_k, \bPsi_k$, however requires special care, and some technical details that we relegate to the appendix. 
The net result of our derivations is the Mixtures of Factor Analysers of Data (MixFAD) algorithm that we have implemented in a R~\citep{R} package of the same name. 


\subsubsection{Choosing the number of mixture components and latent factors}
The developments in the preceding sections have been for a fixed number of mixture components $K$ and a fixed number of latent factors $q$, that we have assumed, for convenience, to be the same for all the mixture components. We use the Bayesian information criterion (BIC) of \citet{schwarz1978}: $-2\hat{l}+\nu\log(n)$, where $\hat{l}$ is the observed data loglikelihood computed from the fitted model, and $\nu$ is the total number of model parameters. Since a smaller BIC value indicates a larger loglikelihood and lower penalty terms relate to less complicated models, and hence a better fit of the model to the data, we choose the $(K,q)$-pair minimising the BIC. From these optimal values, we obtain the corresponding parameter estimates and then assign each observation a group membership.

In this section, we have provided methodology that provides for ellipsoidally-shaped clusters while also characterising the variability in these clusters by means of a few group-wise latent factors. We illustrate our use of this methodology to clarify ellipsoidally-shaped clusters in the thinned completely observed BATSE catalogue GRBs in Section~\ref{sec:app}. We next discuss the issue of finding more complex-featured groups to explain the GRBs.

\subsection{Model-estimated Overlap-Based Syncytial Clustering}
\label{sec:merge}
Clustering algorithms such as MBC are built on the premise that there is a one-to-one correspondence between a mixture component and group. (In the case of GMM, for instance, this gives rise to ellipsoidally-shaped groups.) Such assumptions may not always be reasonable, and indeed, some of these clusters may actually be subgroups of larger more generally-structured groups, so some authors~\citep{baudryetal10,hennig10,melnykov16,petersonetal18,almodovarandmaitra20} developed what \citet{almodovarandmaitra20} christened as syncytial clustering which merges poorer-separated groups to form larger super-clusters. \citet{almodovarandmaitra20} developed a data-driven algorithm that merged groups from the output of a $k$-means clustering algorithm. Subsequent development~\citep{chattopadhyayetal22} extended it to MBC, that we also adopt and adapt in this paper for use in categorising GRBs. Our approach first uses MixFAD  to obtain Gaussian-distributed clusters with factor-structured covariance matrices, or {\em simple} groups in the definition of \citet{almodovarandmaitra20}, and then successively merges the poorer-separated groups obtained from MixFAD, using pairwise and generalised overlaps~\citep{maitraandmelnykov10,melnykovandmaitra11,melnykovetal12}, to obtain {\em compound} or {\em composite} groups, in an algorithm that we outline here. 


\subsubsection{Overlap Measures between Simple Clusters}
\label{sec:overlap-simple}
Suppose that the initial clustering produces $K$ simple groups. Then, the probability that a data vector $\bx_i|z_i=k_1$ is misclassified to the $k_2$th cluster is given by \citet{maitraandmelnykov10}
\begin{equation}
    \omega_{k_2|k_1} = \Pr\left(
    \frac{\eta_{k_2}\bphi(\bx_i;\bmu_{k_2},\bSigma_{k_2})}
    {\eta_{k_1}\bphi(\bx_i;\bmu_{k_1},\bSigma_{k_1})}
    >1\right).
\end{equation}
Then, the pairwise overlap between the $k_1$th and $k_2$th clusters is $\omega_{k_1,k_2} = \omega_{k_1|k_2}+\omega_{k_2|k_1}$ \citep{maitraandmelnykov10}. The pairwise overlap measure ranges from 0 to 1, and  reveals the distinctiveness between two estimated clusters. Specifically, a smaller  overlap value indicates greater separation between clusters, and the converse is also true with an overlap of unity meaning no distinction between the two groups. Further, we can obtain a generalised overlap $\ddot\omega$ from the $K\times K$ matrix $\bOmega$ of pairwise overlaps as $\ddot\omega = (\lambda^*_{\bOmega}-1)/(K-1)$, where $\lambda^*_{\bOmega}$ is the largest eigenvalue of $\bOmega$ \citep{melnykovandmaitra11} by borrowing an idea from \citet{maitra10}. The generalised overlap indexes the overall clustering complexity in terms of a single index, with smaller values indicate a larger overall separation between the $K$ groups. Analytical computations of these overlap measures for GMM are implemented in the R~\citep{R} \textit{MixSim} package \citep{melnykovetal12}.

\subsubsection{Overlap Measures between Compound Clusters}
\label{sec:overlap-compound}
A compound (or composite) cluster is obtained by merging simple clusters, and the probability that a data vector $\bx_i$ from the $m_1$th compound cluster $\mathcal{C}_{m_1}$ is misclassified to the $m_2$th compound cluster $\mathcal{C}_{m_2}$ is given in our case by
\begin{equation}
    \omega_{m_2|m_1} = \Pr\left(
    \frac{\sum_{j\in\mathcal{C}_{m_2}}\eta_{j}\bphi(\bx_i;\bmu_{j},\bSigma_{j})}
    {\sum_{l\in\mathcal{C}_{m_1}}\eta_{l}\bphi(\bx_i;\bmu_{l},\bSigma_{l})}
    >1\right),
\end{equation}
which can, as in \citet{chattopadhyayetal22}, be approximated via Monte Carlo methods as follows:
\begin{enumerate}
    \item Generate M (set to be 100,000 in Section \ref{sec:app}) samples of $\bx_i, i = 1,2,\cdots,M$ from the mixture distribution of the $m_1$th compound cluster $\mathcal{C}_{m_1}$, which is defined as $\sum_{l\in\mathcal{C}_{m_1}}\eta^*_{l}\phi(\cdot;\bmu_{l},\bSigma_{l})$ with $\eta^*_{l} = \eta_{l}/\sum_{h\in\mathcal{C}_{m_1}}\eta_{h}$.
    \item Compute the pairwise overlap between the two compound clusters $\mathcal{C}_{m_1}$ and $\mathcal{C}_{m_2}$ as $\omega_{m_1,m_2} = \omega_{m_1|m_2}+\omega_{m_2|m_1}$ where we estimate 
    \begin{equation}
       \hat\omega_{m_2|m_1} = \displaystyle\frac{1}{M}\sum_{i=1}^{M}\boldone\left\{\frac{\sum_{j\in\mathcal{C}_{m_2}}\eta_{j}\bphi(\bx_i;\bmu_{j},\bSigma_{j})}
    {\sum_{l\in\mathcal{C}_{m_1}}\eta_{l}\bphi(\bx_i;\bmu_{l},\bSigma_{l})}>1\right\}.
    \end{equation}   
\end{enumerate}
Accordingly, the generalised overlap can be computed using these pairwise overlaps, in a similar manner as described in Section \ref{sec:overlap-simple}.

\subsubsection{The MOBSynC Procedure}
\label{sec:procedure}
We now provide the specific MOBSynC procedure for merging the MBC-obtained $K$ groups:
\begin{enumerate}
    \item For the $K(K-1)/2$ pairs of simple clusters, compute the pairwise overlaps $\omega_{k_1,k_2}$ and the generalised overlap $\ddot\omega$ according to Section \ref{sec:overlap-simple}.
    \item Merge the $k_1$th and $k_2$th clusters if $\omega_{k_1,k_2}>\kappa\ddot\omega$, where the parameter $\kappa$ indicates  merging reluctance, with a larger value indicating that fewer pairs are merged in this step. Selection of $\kappa$ is discussed in Remark \ref{remark:kappa}.
    \item For the compound clusters from the merging step (ii), compute all the pairwise overlaps $\omega_{k_1,k_2}$ and the generalised overlap $\ddot\omega$ according to Section \ref{sec:overlap-compound} and then repeat step (ii).
    \item Stop merging if the generalised overlap at current stage, or its change compared to the previous stage, is less than $\epsilon$, set in generally to be a small value, and here to be $10^{-3}$.
\end{enumerate}

\begin{remark}
    \label{remark:kappa}
        \citet{almodovarandmaitra20} provide a data-driven method to determine an integer value of $\kappa$ by running the merging procedure with $\kappa = 1, 2,\ldots,\kappa_{max}$, where $\omega_{k_1,k_2}<\kappa_{max}\ddot\omega$ for all the paired clusters. The optimal $\kappa$ is selected as the one produces the smallest generalised overlap for the final partitioning. 
    \end{remark}
    \begin{remark}
	    \label{remark:stop}
	    Step (iv) in the algorithm was meant by \citet{almodovarandmaitra20} to determine the number of general-structured groups. In our context, we are not particularly interested in the number of general-shaped groups, but rather in the use of MOBSynC to describe  the hierarchical sub-group structure of the BATSE GRBs to provide greater context to the ellipsoidally-shaped groups repeatedly found in several recent studies.
\end{remark}

\begin{figure}
 \centering
\subfloat[True clusters]{\includegraphics[width=0.24\textwidth]{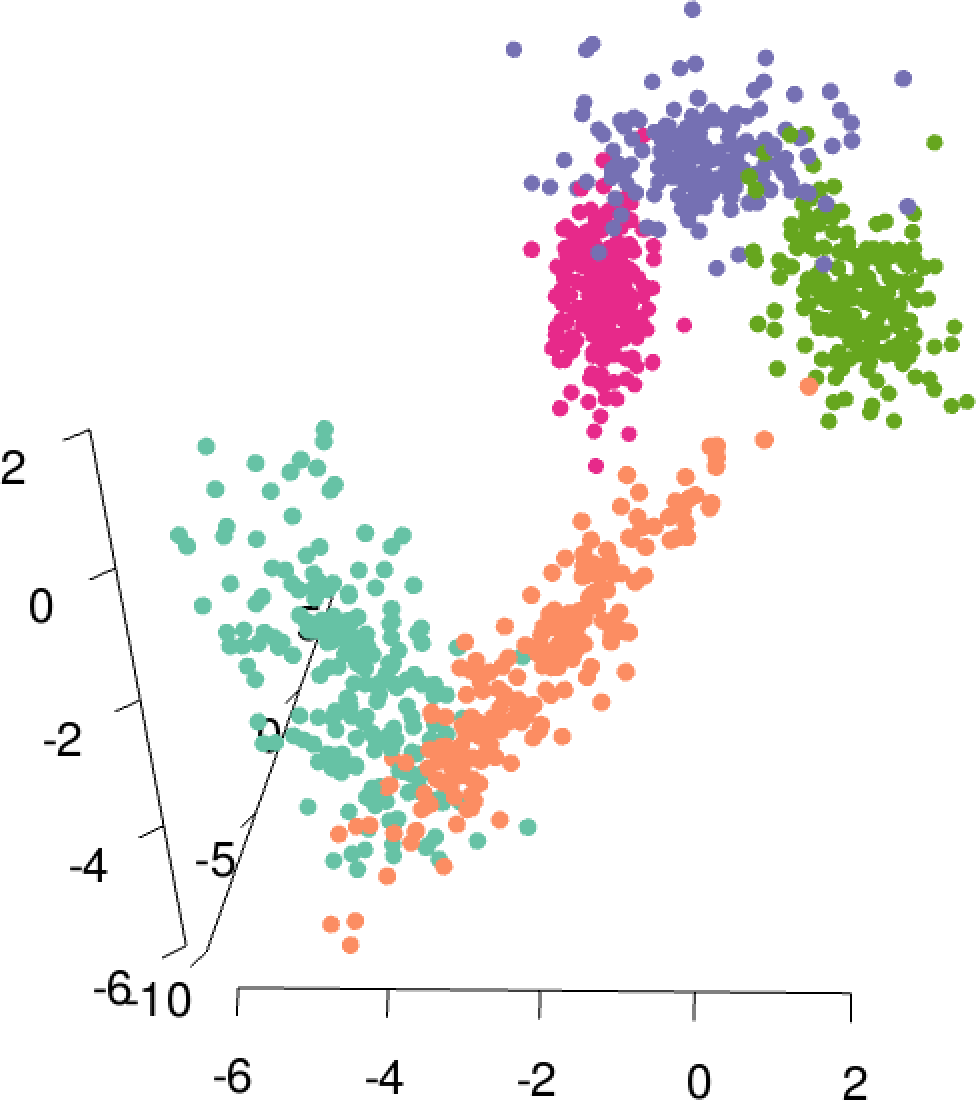}\label{fig:sim-5g-simple}}%
\subfloat[Estimated clusters]{\includegraphics[width=0.24\textwidth]{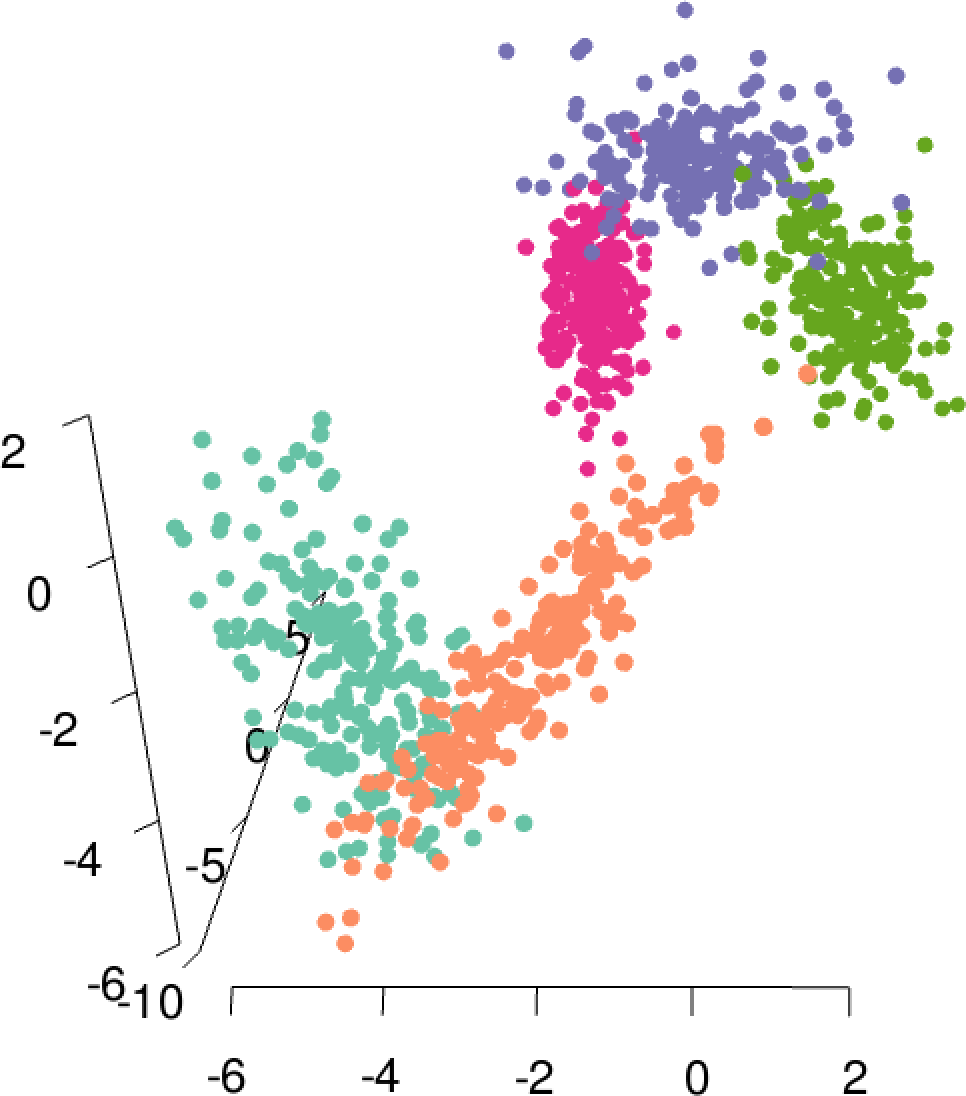}\label{fig:sim-5g-fit}}\\ 
\vspace{0.1in} 
 \text{
     Simple Cluster \tikz\draw[sim1,fill=sim1] (0,0) circle (.5ex); 1
     \tikz\draw[sim2,fill=sim2] (0,0) circle (.5ex); 2
     \tikz\draw[sim3,fill=sim3] (0,0) circle (.5ex); 3
     \tikz\draw[sim4,fill=sim4] (0,0) circle (.5ex); 4
     \tikz\draw[sim5,fill=sim5] (0,0) circle (.5ex); 5
    }\\
\subfloat[$\ddot\omega = 0.011$]{\includegraphics[width=0.22\textwidth]{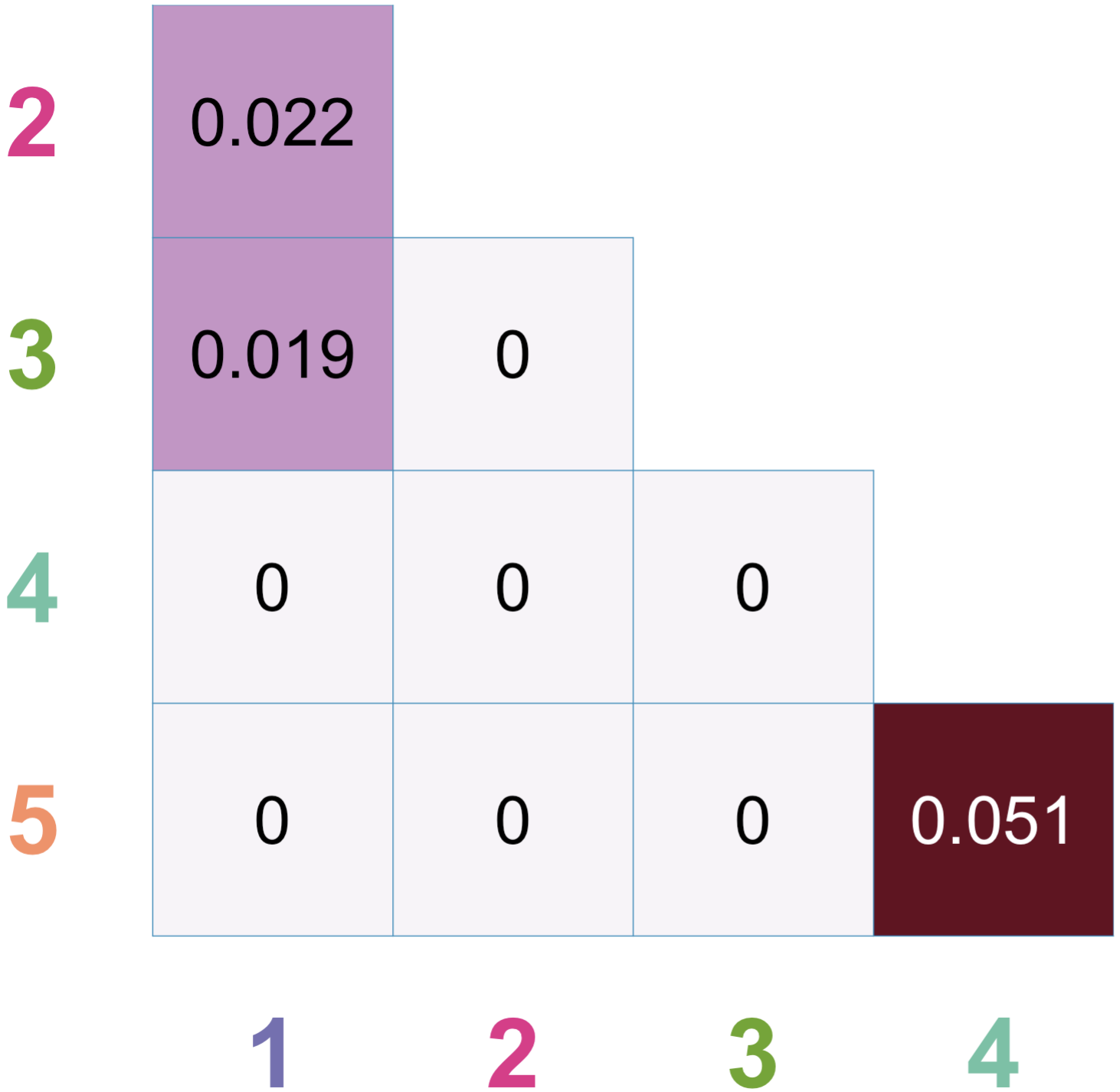}\label{fig:sim-5g-overlap}}%
\subfloat[MOBSynC clusters]{\includegraphics[width=0.26\textwidth]{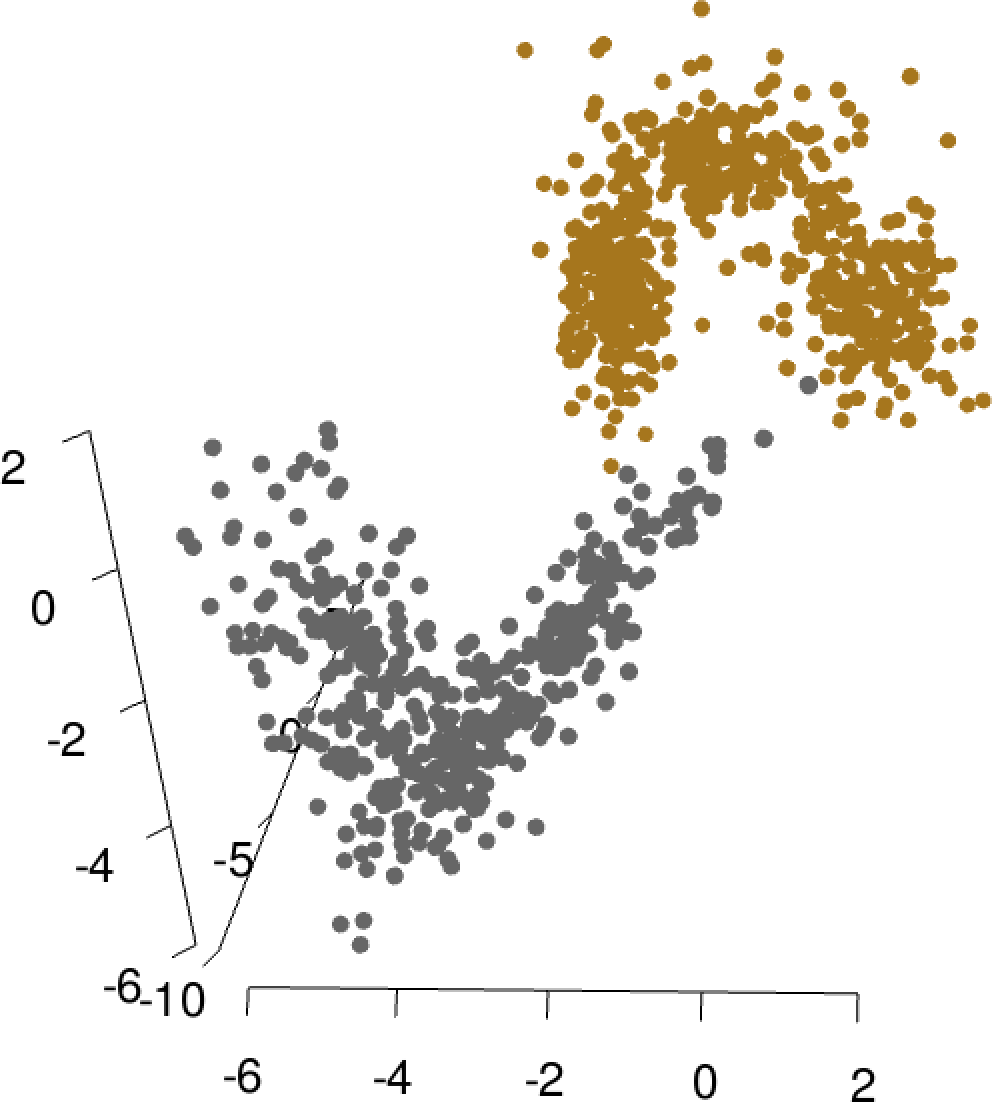}\label{fig:sim-5g-mobsync}}\\
    \vspace{0.1in} 
    \text{
     Composite Cluster \tikz\draw[comp1,fill=comp1] (0,0) circle (.5ex); (1, 2, 3)
     \tikz\draw[comp2,fill=comp2] (0,0) circle (.5ex); (4, 5)
    }\\
    \caption{Simulated dataset: (a) 3D scatter plot showing the true clustering structure, with each different color representing ellipsoidally-shaped Gaussian clusters and the lighter and darker colors representing a complex-shaped ``V'' and a similar ``U'' cluster, (b) the estimated clusters from GMMBC that finds five ellipsoidally-shaped groups with (c) pairwise overlaps, and (d) the composite clusters from MOBSynC recolored to emphasize the composite and compound structure elicited by the algorithm.}
    \label{fig:sim-5g}
\end{figure}
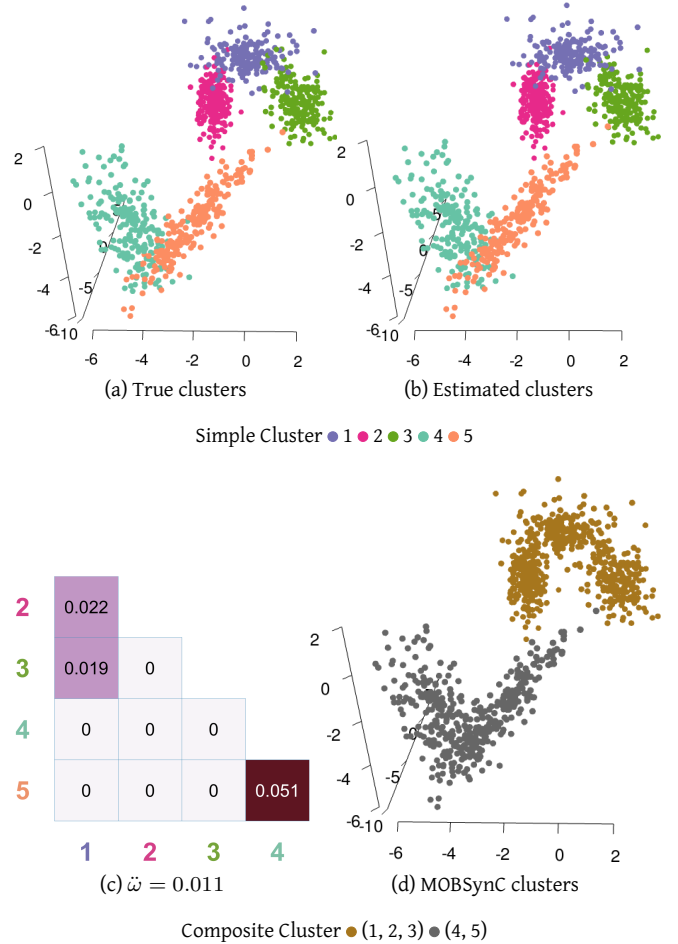

\paragraph{An Illustrative Example}
\label{sec:illustration}
We  illustrate MOBSynC by a simulated example, where a 3D dataset of $n=1000$ observations was randomly generated from a five-component GMM with two separate compound groups that are shaped as  a ``U'' and a ``V'', as shown by means of the darker and lighter colors in Figure~\ref{fig:sim-5g-simple}. Each of the parts of the ``U'' and the ``V'' can be fit by individual ellipsoidal clusters, and these five clusters are described by means of color (whether light or dark) in Figure~\ref{fig:sim-5g-simple}. We fit the data using a GMM of up to $9$ components, and obtained the optimal number of components of $K=5$ by BIC. The fitted results are shown in Figure~\ref{fig:sim-5g-fit} where the five individual, and ellipsoidal, clusters were captured by the GMM: these five simple clusters had 19 observations that were incorrectly misclassified, yielding a classification error rate of 0.019.

We used MOBSynC to investigate if the five ellipsoidal (Gaussian) groups are part of some compound clusters. Figure~\ref{fig:sim-5g-overlap} displays the pairwise overlap map calculated between the 
five simple clusters obtained from fitting the GMM to the data: the generalised overlap was calculated to be $\ddot\omega = 0.011$. By Remark~\ref{remark:kappa}, we determined $\kappa=1$ and obtained $\omega_{1,2},\omega_{1,3},\omega_{4,5}>\kappa\ddot\omega$, indicating the merging of clusters 1, 2 and 3, and of clusters 4 and 5. Then, we computed the overlap measures for the compound clusters (1, 2, 3) and (4, 5), which produced a generalised overlap $\ddot\omega <0.001$, so we stopped merging at the current stage, and finally obtained two composite clusters shown in Figure~\ref{fig:sim-5g-mobsync}, where we map the composite clusters for greater clarity. We see that the merging procedure is able to describe the multi-layered grouping characteristics of the original dataset faithfully.

In this section, we have outlined our needed methodology for MixFAD and MOBSynC. We now apply it to the observed GRBs in the thinned complete BATSE catalogue.

\section{Statistical Analysis of GRBs}
\label{sec:app}
\subsection{Simple ellipsoidal cluster analysis using MixFAD}
\label{sec:grb-gmm}
\begin{figure}
\centering
 \includegraphics[width = 0.8\columnwidth]{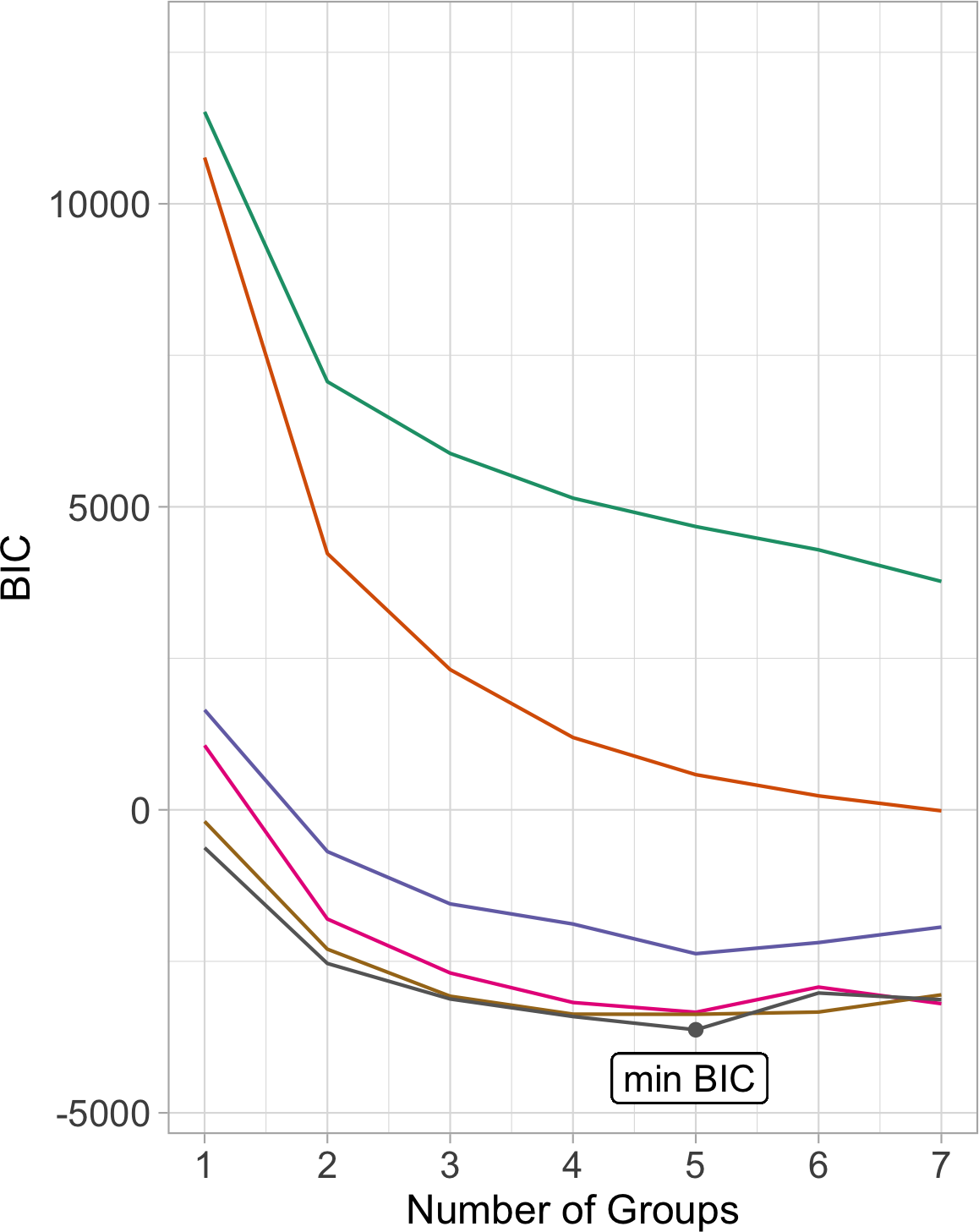}
    \text{
     Number of factors \tikz\draw[err1,fill=err1] (0,0) circle (.5ex); 0
     \tikz\draw[err2,fill=err2] (0,0) circle (.5ex); 1
     \tikz\draw[err3,fill=err3] (0,0) circle (.5ex); 2
     \tikz\draw[err4,fill=err4] (0,0) circle (.5ex); 3
     \tikz\draw[err5,fill=err5] (0,0) circle (.5ex); 4
    \tikz\draw[err6,fill=err6] (0,0) circle (.5ex); 5
    
    }
    \caption{BIC for each $K$ and $q$ upon fitting the Gaussian mixture of factor analysers with the 1150 thinned complete BATSE Catalog GRBs. The minimum BIC is attained at $(K,q) = (5,5)$.}
    \label{fig:grb-bic}
\end{figure}

In the initial clustering phase, the MixFAD algorithm  was fit to the 1150 records from the thinned complete BATSE GRB Catalog with up to seven mixture components and up to five factors (the largest $q$ possible to be fit for  when $p=9$). The optimal numbers of clusters and factors were determined by  BIC to be $K=5,q=5$ (see Figure \ref{fig:grb-bic}).

We investigate the properties of the identified groups obtained and explore the astrophysical characteristics
of the GRBs from individual clusters in terms of the nine features. Table \ref{tab:grb-groupsize} gives the sample size for each of the five simple clusters obtained from the Gaussian mixture of factor analysers, where we see that the second cluster includes the largest number of GRBs while the first and the fifth clusters contain the fewest  bursts. 
\begin{table}
\centering
\caption{\label{tab:grb-groupsize} Number of GRBs in each of the five simple clusters.}
\begin{tabular}{c|c|c|c|c|c}
\hline\hline
\bf{Cluster} & 
\bf{\textcolor{err1}{1}} & \bf{\textcolor{err2}{2}} & \bf{\textcolor{err3}{3}} & \bf{\textcolor{err5}{4}} & \bf{\textcolor{err6}{5}}\\ 
    \hline\hline
\bf{Size} & $156$ & $392$ & $235$ & $248$ & $119$ \\
\hline
\end{tabular}
\end{table}
Table \ref{tab:grb-feature} summarises the means and standard deviations of the nine parameters in each of the individual clusters. It can be seen that the two time duration parameters illustrate the major distinctiveness among the clusters, as we see that the two smallest simple groups have lower average  $\log_{10}T_{50}$ and $\log_{10}T_{90}$ compared to the other three simple  clusters. A similar pattern is also evident for the four time-integrated fluence parameters. 
For the peak fluxes measured in the three bins, we find that our fourth group notably has the highest average level, while the third cluster yields the lowest level on the average. On the other hand, for $\log_{10}P_{1024}$, the fifth cluster also contains small averaged values even though its peak fluxes in bins of 64 and 256 remain relatively high. It is these distinctions that provide definition of the five simple groups found by the MixFAD algorithm in the thinned complete BATSE Catalog.
\begin{table*}
\centering
\caption{Estimated group-wise means and standard deviations (in parenthesis) for each of the nine parameters (after $\log_{10}$ transformation).}
\label{tab:grb-feature} 
\addtolength{\tabcolsep}{-0.4em}
\begin{tabular}{c|c|c|c|c|c|c|c|c|c}
\hline\hline
\diagbox[width=7em]{\bf{Cluster}}{\bf{Feature}} 
&\bm{$\log_{10}T_{50}$}& \bm{$\log_{10}T_{90}$} & \bm{$\log_{10}F_1$}& \bm{$\log_{10}F_2$}& \bm{$\log_{10}F_3$}& \bm{$\log_{10}F_4$}& \bm{$\log_{10}P_{64}$}& \bm{$\log_{10}P_{256}$}& \bm{$\log_{10}P_{1024}$}\\
\hline\hline
\bf{\textcolor{err1}{1}} 
& $-0.75 (0.36)$ &$-0.39 (0.35)$ &$-7.84 (0.38)$ &$-7.58 (0.38)$ &$-6.74 (0.34)$ &$-6.21 (0.51)$  &$0.53 (0.33)$  &$0.35 (0.31)$ &$-0.07 (0.33)$ 
\\
\hline
\bf{\textcolor{err2}{2}} 
& $1.17 (0.45)$  &$1.64 (0.39)$ &$-6.11 (0.39)$ &$-5.99 (0.39)$ &$-5.55 (0.41)$ &$-5.46 (0.52)$  &$0.29 (0.24)$  &$0.24 (0.26)$  &$0.19 (0.27)$
\\
\hline
\bf{\textcolor{err3}{3}} 
& $1.14(0.56)$  &$1.49(0.50)$ &$-6.69(0.54)$ &$-6.60(0.48)$ &$-6.17(0.46)$ &$-5.68(0.44)$  &$0.04(0.24)$ &$-0.07(0.26)$ &$-0.19(0.26)$
\\
\hline
\bf{\textcolor{err5}{4}} 
&$0.87(0.48)$ &$1.43(0.49)$ &$-5.84(0.53)$ &$-5.68(0.52)$ &$-5.16(0.57)$ &$-5.00(0.77)$  &$0.93(0.40)$ &$0.89(0.41)$ &$0.79(0.42)$ 
\\
\hline
\bf{\textcolor{err6}{5}} 
&$-0.08(0.68)$ &$0.46(0.69)$ &$-7.32(0.68)$ &$-7.14(0.66)$ &$-6.44(0.62)$ &$-5.89(0.63)$  &$0.56(0.43)$  &$0.39(0.45)$ &$0.07(0.48)$ 
\\ 
\hline
\end{tabular}
\end{table*}

\begin{table*}
\centering
\caption{Estimated factor loadings (in the correlation scale), along with a heatmap for ready and easy reference. For clarity
of presentation, values in the interval (-0.1,0.1) are suppressed in the table, but displayed using light colors in the heatmap.} 
\label{tab:grb-loadings}
\addtolength{\tabcolsep}{-0.5em}
\begin{tabular}{c|c|c|c|c|c|c|c|c|c|c|c}
\hline
 \hline
\bf{$\bk$}& \bf{$\bq$}& \bm{$\log_{10}T_{50}$}& \bm{$\log_{10}T_{90}$} & \bm{$\log_{10}F_1$}& \bm{$\log_{10}F_2$}& \bm{$\log_{10}F_3$}& \bm{$\log_{10}F_4$}& \bm{$\log_{10}P_{64}$}& \bm{$\log_{10}P_{256}$}& \bm{$\log_{10}P_{1024}$}&\bf{Heatmap representation}\\
  \hline
 \hline
\multirow{ 5}{*}{\bf{\textcolor{err1}{1}}} & 1 
 &
 &
 &
 &
 &0.183
 &
 &0.615 
 &0.970
 &0.187 
 &
  \multirow{ 25}{*}
 {
 \includegraphics[height=9.25cm]{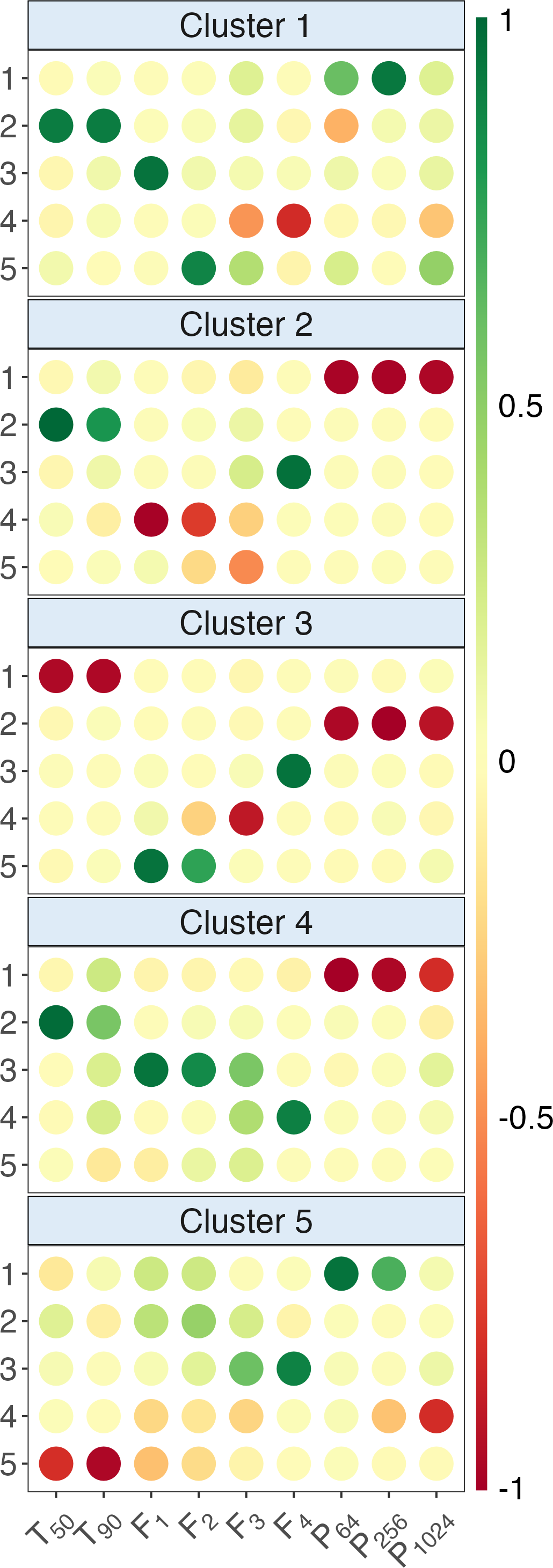}
    }
 \\
  \textbf{}& 2 
 &0.952
 &0.953
 &
 &
 &0.144
 &
 &-0.381 
 &
 &0.111
 & \\
 & 3 
 &
 &
 &0.992
 &
 &
 &
 &
 &
 &0.129 
 & \\
  & 4 
 &
 &
 &
 &
 &-0.481
 &-0.834
 &
 &
 &-0.305
 & \\
  & 5 
 &
 &
 &
 &0.916
 &0.370
 &
 &0.218 
 &
 &0.479
 &\\
 \cline{1-11}
 \multirow{ 5}{*}{\bf{\textcolor{err2}{2}}} & 1 
 &
 &
 &
 &
 &-0.124
 &
 &-0.991 
 &-0.992
 &-0.988 &\\
  & 2 
 &0.996
 &0.842
 &
 &
 &0.112
 &
 &
 &
 &
 & \\
 & 3 
 &
 &
 &
 &
 &0.222
 &0.991
 &
 &
 &
 & \\
 & 4 
 &
 &
 &-0.993
 &-0.776
 &-0.262
 &
 &
 &
 &
 & \\
 & 5 
 &
 &
 &
 &-0.217
 &-0.516
 &
 &
 &
 &
 & \\
 \cline{1-11}
 \multirow{ 5}{*}{\bf{\textcolor{err3}{3}}} & 1 
 &-0.979
 &-0.984
 &
 &
 &
 &
 &
 &
 &
 & \\
 & 2 
 &
 &
 &
 &
 &
 &
 &-0.986 
 &-0.995
 &-0.937
 & \\
 & 3 
 &
 &
 &
 &
 &
 &0.993
 &
 &
 &
 & \\
 & 4 
 &
 &
 &
 &-0.252 
 &-0.914
 &
 &
 &
 &
 & \\
 & 5 
 &
 &
 &0.992
 &0.779
 &
 &
 &
 &
 &
 & \\
 \cline{1-11}
 \multirow{ 5}{*}{\bf{\textcolor{err5}{4}}} & 1 
 &
 &0.267 
 &
 &
 &
 &
 &-0.994
 &-0.985
 &-0.831
 & \\
 & 2 
 &0.994
 &0.566
 &
 &
 &
 &
 &
 &
 &
 & \\
 & 3 
 &
 &0.205 
 &0.986 
 &0.891
 &0.560
 &
 &
 &
 &0.159
 & \\
 & 4 
 &
 &0.224
 &
 &
 &0.379
 &0.936
 &
 &
 &
 & \\
 & 5 
 &
 &-0.138
 &-0.105 
 &0.122
 &0.194
 &
 &
 &
 &
 & \\
 \cline{1-11}
 \multirow{ 5}{*}{\bf{\textcolor{err6}{5}}} & 1 
 &-0.139
 &
 &0.268 
 &0.264
 &
 &
 &0.995
 &0.702
 &
 & \\
 & 2 
 &0.178
 &
 &0.336 
 &0.471
 &0.225
 &
 &
 &
 &
 & \\
 & 3 
 &
 &
 &
 &0.161 
 &0.607
 &0.926 
 &
 &
 &0.105
 & \\
 & 4 
 &
 &
 &-0.229 
 &-0.150
 &-0.238
 &
 &
 &-0.314
 &-0.830
 & \\
 & 5 
 &-0.822
 &-0.987
 &-0.327 
 &-0.208
 &
 &
 &
 &
 &
 & \\
 \cline{1-11}
\end{tabular}
\end{table*}

\subsubsection*{Characterising latent structure in the simple GRB groups}
\label{sec:grb-fa}
As exhibited for GRB data in \citet{bagolyetal09}, a major strength of factor analysis is that it allows for the interpretation of the variability in a dataset in terms of a few linear combinations of the parameters. \citet{bagolyetal09} performed their analysis only on 197 BATSE GRBs of longer duration. Our approach uses MixFAD to perform an integrated characterisation of the variability structure in the obtained groupings while performing cluster analysis, which is a major strength. We now discuss the results that are obtained by MixFAD and that can help in characterising the simple groups beyond that provided by the means and standard deviations that have already been discussed in Table~\ref{tab:grb-feature}, reiterating for clarity that all parameters are discussed here after their $\log_{10}$ transformations. We reiterate here that because of the invariance to orthogonal transformations discussed shortly after \eqref{eq:gmmfad}, the sign of a factor loading does not matter, however, the parity or disparity in sign between two parameter contributions to a factor loading does.

For the estimated factor model, Table \ref{tab:grb-loadings} displays both numerically and also,  for ready reference, visually, the fitted factor loadings with oblimin rotation \citep{costello2005} applied to simplify interpretation for the five simple clusters. These factor loadings are the five columns of the $k$th group's  $p\times q$ matrix $\bLambda_k$ in \eqref{eq:gmmfad}. The proportion of the total variances explained by the five factors are $74.6\%$, $87.9\%$, $92.5\%$ $83.2\%$ and $67.6\%$, in that order for the five clusters numbered 1 through 5, indicating that the factors fairly adequately summarise the variability in each of the estimated simple GRB groups. Further, we list only those numerical values in the loadings on Table~\ref{tab:grb-loadings} that are not negligible, in the sense that these values all are outside the interval $(-0.1,0.1)$.  We now discuss and distinguish the obtained five factor loadings individually for each simple group.

For the first cluster, we see that the first factor is a weighted average of $F_3$ and the three peak fluxes with small to substantial to large contributions from $F_3$ and $P_{1024}$, $P_{64}$, and $P_{256}$ in that order, while the second factor is explained mainly by the two flux arrival times along with minor amounts of $F_3$ and $P_{1024}$, and small contribution from $P_{64}$ in the opposite direction (negative sign). We see that the third factor loading vector is mainly dominated by the time-integrated fluence in the 20-50 keV ($F_1$), and a smattering of $P_{1024}$. The fourth factor loading vector is a linear combination of small to moderate to substantial contributions from $P_{1024}$, $F_3$, and $F_4$ in that order. Finally, the last factor for this group is majorly determined by $F_2$ along with moderate contributions from $P_{1024}$ and small parts of $F_3$ and $P_{64}$. 

The first factor in the second cluster is almost entirely a simple average of the three peak fluxes along with a minor amount of $F_3$, while the second factor is essentially a weighted average of two flux arrival times, with $T_{50}$ a bit more dominant than $T_{90}$, and a minor contribution from $F_3$. For this second group, the third factor loading vector is essentially $F_4$ along with a small part of $F_3$, while the fourth factor is a weighted average of the first three time-integrated fluences with very large to substantial to small contributions. Finally, the fifth factor for this cluster is a weighted average of $F_2$ and $F_3$ with the third kind of time-integrated fluence contributing over twice in magnitude to that of the other.

The third cluster has its first factor loading vector as a weighted average dominated by the two flux arrival times. The second loading vector is again a weighted average, predominantly of the three peak fluxes. The third factor is essentially $F_4$. The fourth factor is a weighted average, of very high to small contributions from $F_3$ and $F_2$ in that order. Finally, the fifth factor loading vector is a weighted average of $F_1$ and $F_2$, with the former contributing about $30\%$ more than the latter.

Our fourth group has the first factor loading vector as a contrast between a small amount of $T_{90}$ and a weighted average of the three peak fluxes (with very high contributions from $P_{64}$, $P_{256}$ and lesser from $P_{1024}$). The second factor loading vector is mainly dominated by the two flux arrival times with $T_{50}$ contributing almost twice to that of $T_{90}$, while the third factor is a weighted average of a small part of $T_{90}$ and the first three time-integrated fluences (with very high to substantial to moderate contributions). The fourth factor loading vector is majorly explained by $F_4$, and small amounts of $T_{90}$ and $F_3$. Finally, the fifth factor is a contrast between a weighted average of $T_{90}$ and $F_1$ on one hand, and a weighted average of $F_2$ and $F_3$ from the other side (all with fairly minor contributions however).

Our fifth and final simple cluster has the first loading vector as a contrast between a minor amount of $T_{50}$, and a weighted average with small, substantial and very large contributions from $F_1$ and $F_2$, $P_{256}$ and $P_{64}$ in that order from the opposite direction, while the second factor is a linear combination of a small amount of $T_{50}$ and the first three time-integrated fluences with $F_2$ contributing more than $F_1$ and $F_3$. The third factor is a weighted average of a minor amount of $P_{1024}$ and the last three time-integrated fluences with low to moderate to very high contributions. Further, the fourth factor is again a weighted average of a large contribution from $P_{1024}$ and small amounts of $P_{64}$ and the first three time-integrated fluences. Finally, the fifth factor is mainly explained by the two flux arrival times with $T_{90}$ a bit more dominant than $T_{50}$ along with small contributions from $F_1$ and $F_2$. In summary, the factor loadings for the five simple groups are collectively distinct, and provide additional definition to the dispersion of the GRBs in each of them. 

We also estimated the factor scores $\bF_{ik}$ (described in Section \ref{sec:fa}) to further characterise the simple clusters. We use the commonly-applied method of \cite{thurstone35}, where, for $\bx_i$ that is classified to be in the $k$th simple group, we minimise the weighted squared error loss $\|\bPsi_k^{-1/2}(\bx_i - \bmu_k - \bLambda_k\bF_{ik})\|^2$. The corresponding solution is the  \citep{bartlett37} score that is given by 
$\widehat{\bF}_{ik} = (\bLambda_k^\top\bPsi_k^{-1}\bLambda_k)^{-1}\bLambda_k^\top
  \bPsi_k^{-1}(\bx_i - \bmu_k)$ and that unbiasedly estimates the true factor score \citep{hershberger05,distefanoetal09}.
\begin{table}
\centering
\caption{The mean factor scores for the five simple clusters.}
\label{tab:grb-5g-fascore} 
\addtolength{\tabcolsep}{-0.6em}
\begin{tabular}{c|c|c|c|c|c}
\hline\hline
\bf{Cluster}
&\bf{Factor 1}&\bf{Factor 2}& \bf{Factor 3}& \bf{Factor 4}& \bf{Factor 5}\\
\hline\hline
\bf{\textcolor{err1}{1}} 
& 0.008   &0.012   &0.007   &0.004  &-0.002 \\
\bf{\textcolor{err2}{2}} 
&0.019  &-0.003   &0.017   &0.023  &-0.034\\
\bf{\textcolor{err3}{3}} 
&0.091  &0.021   &0.023   &0.064  &-0.053   \\
\bf{\textcolor{err5}{4}} 
&-0.086  &-0.097  &-0.059  &-0.015  &-0.009  \\
\bf{\textcolor{err6}{5}} 
&-0.069  &-0.102  &-0.008   &0.071   &0.001    
\\
\hline
\end{tabular}
\end{table}
Consequently, the factor score of $\bx_i$ represents the $i$th observations numerical weight, or its "ratings", of the importance on the $q$ latent factors in its characterisation~\citep{distefanoetal09}.
Table~\ref{tab:grb-5g-fascore} lists the averaged factor scores for the five simple clusters, where we find that for the first simple group, the average importance of the underlying factors as "rated" by its GRB members, is in the order of factor 2 $>$ factor 1 $>$ factor 3 $>$ factor 4 $>$ factor 5, while for the second group, these importance ratings are in the order of factor 4 $>$ factor 1 $>$ factor 3 $>$ factor 2 $>$ factor 5. Similarly, the averaged score of the importance of the factors in the third group yields the ordering of factor 1 $>$ factor 4 $>$ factor 3 $>$ factor 2 $>$ factor 5, while that of the fourth group is factor 5 $>$ factor 4 $>$ factor 3 $>$ factor 1 $>$ factor 2, and for the fifth group, its GRBs "rate" important factors to be in the order of factor 4 $>$ factor 5 $>$ factor 3 $>$ factor 1 $>$ factor 2. In this analysis also, we see that the average factor scores outline their importance to the groups distinctly and differentially across each of the different simple groups. We now explore if the five simple groups of the thinned complete BATSE Catalog GRBs can be explained as being part of some complex super-group structure.

\subsection{Identifying complex structure in the groups}
\label{sec:grb-mobsync}
\begin{figure}
\centering
    \mbox{\subfloat[$\ddot\omega = 0.049$\label{fig:grb-1150-overlap-initial}]{\includegraphics[width=0.45\columnwidth]{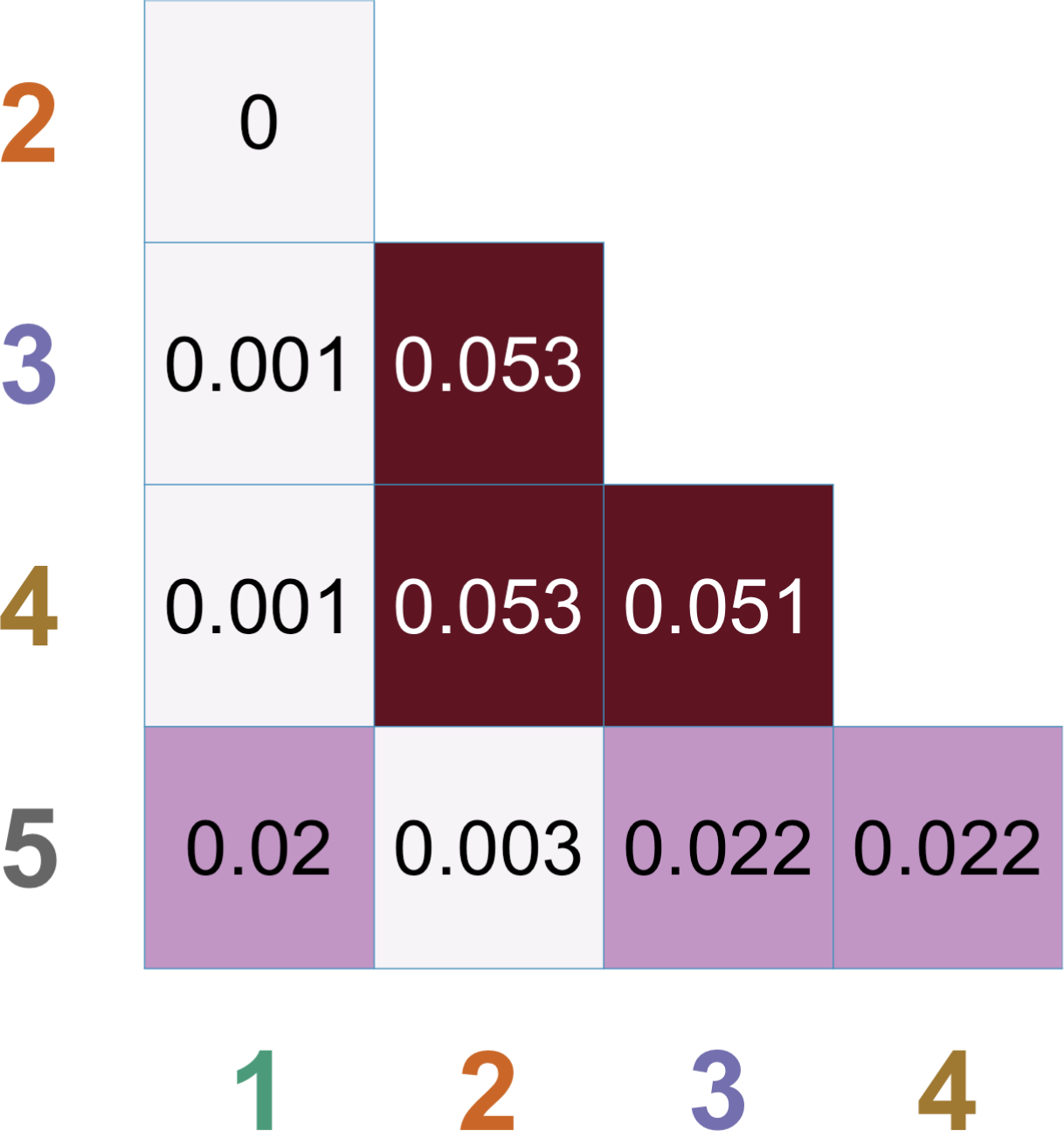}
    }
    \hspace{0.2in}
    \subfloat[$\ddot\omega = 0.042$\label{fig:grb-1150-overlap-compound}]{\includegraphics[width=0.37\columnwidth]{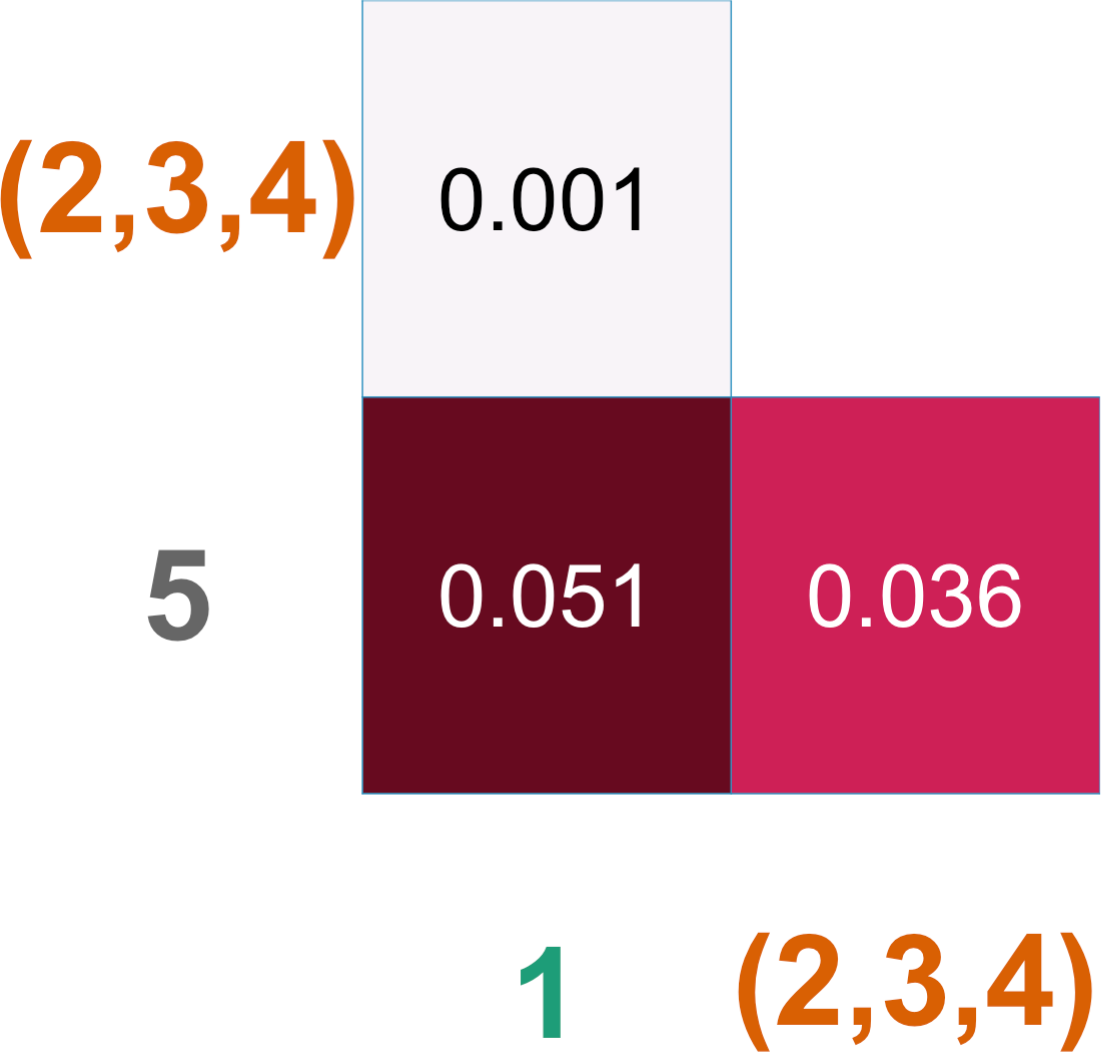}%
    }
}
\caption{The pairwise and generalised overlap measures between (a) the simple clusters at the end of the MixFAD algorithm, and (b) between the groups at the end of the first stage of merging.}
\end{figure}

We applied our MOBSynC algorithm of Section~\ref{sec:merge} to investigate the propensity of merging, and the presence of higher order compound or composite clusters. Figure \ref{fig:grb-1150-overlap-initial} shows the pairwise overlaps between any two clusters and the generalised overlap $\ddot\omega$ of $0.049$. For the selected $\kappa=1$, clusters 2, 3 and 4 merge into one group. For clusters after the first merging, Figure \ref{fig:grb-1150-overlap-compound} presents the pairwise overlaps and the generalised overlap $\ddot\omega$ of $0.042$, which indicates a further merging between the simple groups 1 and 5. Finally, we obtain two compound groups (simple clusters 1,5 v.s. simple clusters 2,3,4) that gives a generalised overlap $\ddot\omega$ of $0.032$.
Our MOBSynC procedure therefore indicates that the thinned complete BATSE Catalog GRBs are grouped at the highest resolution level in terms of five ellipsoidal clusters, and at the intermediate level in terms of three clusters, one of which is a compound group composed of the erstwhile ellipsoidal clusters 2, 3 and 4 and the rest being the two other ellipsoidal clusters. At the final level, we have two compound groups, composed of two and three ellipsoidal groups. We see that our MOBSynC characterisation allows for the elicitation of the complex structures and features in the GRBs. We now analyse the compound groups.

\subsubsection*{Characterising latent structure in the compound groups}
Unlike with the simple groups obtained by Section~\ref{sec:grb-gmm} where the latent structures are found as a byproduct of MixFAD, latent structures hidden in the compound groups can only be clarified after some additional development. This is because any compound cluster found by our methodology is not Gaussian, but a mixture of Gaussians whereas a common assumption in factor analysis is that the observations are from at least an ellipsoidally symmetric, and indeed very often, a Gaussian distribution. Our approach will consider the observations assigned to each compound (composite) cluster, Gaussianise the data in them, and then apply factor analysis similar to that done by \citet{bagolyetal09}, but restricted to only the data from the compound cluster being considered. We now discuss a general approach to Gaussianise a dataset using a Gaussian distributional transform (GDT) \citep{zhuetal21} that is specified within the general framework of copula models~\citep{nelsen06} that were introduced to the astrophysics community by \citet{yuanetal18}.

The basic idea behind the GDT, and indeed copula models, stems from the fact that for any univariate random variable (RV) $X$ having a continuous cumulative distribution function (CDF) ${\mathcal F}(x)$, the distribution of ${\mathcal F}(X)$  is standard uniform. Consequently, the distribution of the transformed RV $Y$ is standard normal, where $Y=\varrho(X) \doteq\Phi_{0,1}^{-1}[{\mathcal F}(X)]$, and $\Phi_{\mu,\sigma^2}(\cdot)$ is the CDF of a $\mN(\mu,\sigma^2)$ RV . Therefore, we have Gaussianised the univariate RV $X$ into a standard normal random variable, and the composition function $\varrho(\cdot)=\Phi_{0,1}^{-1}\circ{\mathcal F}(\cdot)$ is a Gaussianising function. In the multivariate context, we apply the Gaussianising transformation marginally to each of the components of the random vector $\bX=(X_1,X_2,\ldots,X_p)$. That is, we obtain $Y_j=\varrho_j(X) \doteq\Phi_{0,1}^{-1}[{\mathcal F_j}(X_j)]$, where $\mathcal{F}_j(\cdot)$ is the marginal CDF of $X_j$. Then the transformed random vector $\bY = (Y_1,Y_2,\ldots,Y_p)$ has standard normal marginals, but the association between the components in $\bY$ is preserved, and indeed, the variance covariance matrix of $\bY$ is a correlation matrix. We perform factor analysis, using the computationally efficient FAD algorithm of~\citet{daietal21}, on this correlation matrix formed by Gaussianising the data assigned to each compound cluster, after noting that it remains to specify the marginal CDF $\mathcal F_j(\cdot)$ of our compound clusters. From the discussion in Section~\ref{sec:overlap-compound}, the CDF of a random vector $\bX$ in the compound cluster $\mathcal C_m$ is given by $\sum_{l\in\mathcal{C}_{m}}\eta^*_{l}\bPhi(\cdot;\bmu_{l},\bSigma_{l})$ with $\eta^*_{l} = \eta_{l}/\sum_{h\in\mathcal{C}_{m}}\eta_{h}$. 
Therefore, the marginal CDF $\mathcal F_j(\cdot) =  \sum_{l\in\mathcal{C}_{m}}\eta^*_{l}\Phi_{\mu_{lj},\sigma^2_{ij}}(\cdot)$ with $\eta^*_{l} = \eta_{l}/\sum_{h\in\mathcal{C}_{m}}\eta_{h}$, where $\mu_{lj}$ is the $j$th component of the vector $\bmu_l$, and $\sigma^{2}_{lj}$ is the $(j,j)$th diagonal entry of $\bSigma_l$. With these definitions in place, we are now in a position to investigate and characterise the latent structure in the compound clusters. However, we also point out here that the linear relationship of the latent factors are in Gaussianised space and, unlike in the case of simple groups, not in the original data space, nevertheless this approach provides us with a summary of the relative strengths of the nine original parameters in the latent factors to explain the variability in the compound groups.

\begin{table*}
\centering
\caption{Estimated factor loadings (in Gaussianised space) for composite clusters (1, 5) and (2, 3, 4). For clarity
of presentation, values in the interval (-0.1,0.1) are not included in the table. As before, a heatmap provides a quick and easy visual reference.} 
\label{tab:grb-comp-loadings}
\addtolength{\tabcolsep}{-0.72em}
\begin{tabular}{c|c|c|c|c|c|c|c|c|c|c|c}
\hline
 \hline
\bf{$\bk$}& \bf{$\bq$}& \bm{$\log_{10}T_{50}$}& \bm{$\log_{10}T_{90}$} & \bm{$\log_{10}F_1$}& \bm{$\log_{10}F_2$}& \bm{$\log_{10}F_3$}& \bm{$\log_{10}F_4$}& \bm{$\log_{10}P_{64}$}& \bm{$\log_{10}P_{256}$}& \bm{$\log_{10}P_{1024}$}
&\bf{Heatmap representation}\\
  \hline
 \hline
\multirow{ 4}{*}{({\bf\textcolor{err1}{1}}, \bf{\textcolor{err6}{5}})}
& 1 
 &0.331
 &0.362
 &0.824 
 &0.877 
 &0.901
 &0.657
 &0.880 
 &0.962
 &0.963 
 & \multirow{ 9}{*}
 {
 \includegraphics[height=3.7cm]{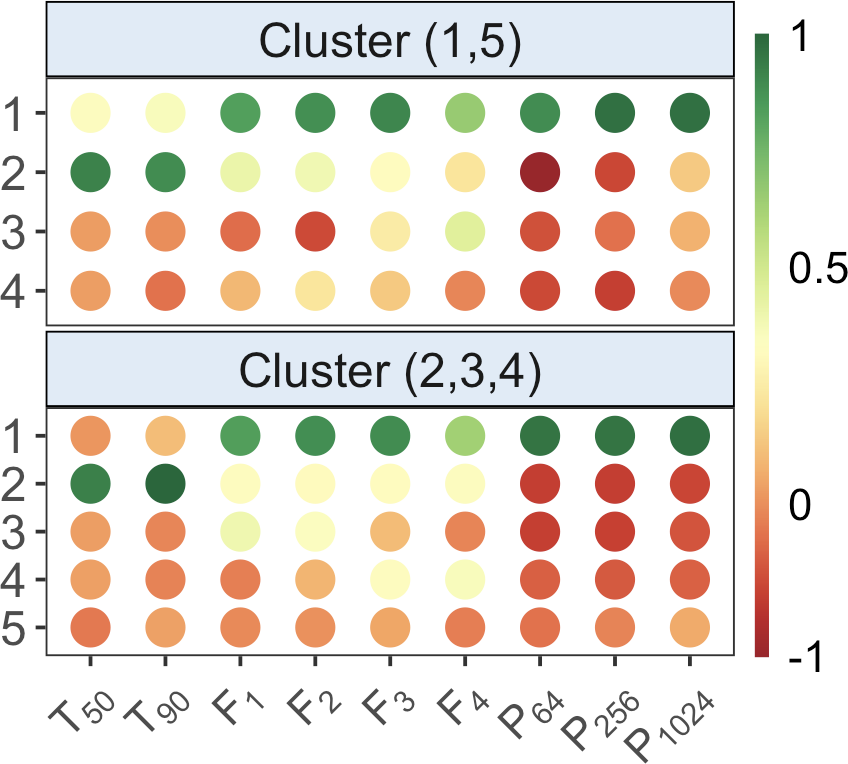}
    }
 \\
  & 2 
 &0.912
 &0.885
 &0.409 
 &0.388 
 &0.320
 &0.219
 &-0.322 
 &-0.172
 &0.140
 &
 \\
 & 3 
 &
 &
 &
 &-0.162 
 &0.248
 &0.450
 &-0.148 
 &
 &
 &
 \\
 & 4 
 &
 &
 &
 &0.223 
 &0.138
 &
 &-0.167
 &-0.191
 &
 &
 \\
 \cline{1-11}
 \multirow{ 5}{*}{({\bf\textcolor{err2}{2}}, {\bf\textcolor{err3}{3}}, \bf{\textcolor{err5}{4}})} 
 & 1 
 &
 &0.106
 &0.827
 &0.879
 &0.882 
 &0.630
 &0.951
 &0.952 
 &0.965
 &
 \\
  \textbf{}& 2 
 &0.917
 &0.990
 &0.323 
 &0.312 
 &0.316
 &0.331
 &-0.196
 &-0.197
 &-0.175
 &
 \\
 & 3 
 &
 &
 &0.395 
 &0.342 
 &0.102
 &
 &-0.193 
 &-0.186
 &-0.136 
 &
 \\
  & 4 
 &
 &
 &
 &
 &0.322
 &0.358
 &-0.108
 &-0.123
 &-0.105
 &
 \\
 & 5 
 &
 &
 &
 &
 &
 &
 &-0.100
 &
 &0.100 
 &
 \\
 \cline{1-11}
\end{tabular}
\vspace{0.1in}
\end{table*}

Using the above approach, we summarise the sole compound or composite cluster (2, 3, 4) formed after the first stage of merging. (The remaining clusters 1 and 5, and therefore their latent characterisations, are unchanged from Section~\ref{sec:grb-gmm}.) For the GRBs assigned to this cluster, the average levels of $\log_{10}T_{50}$ (1.076), $\log_{10}T_{90}$ (1.537) and $\log_{10}P_{1024}$ (0.258) are notably higher compared to the other
two simple clusters 1 and 5 (mean values in Table~\ref{tab:grb-feature}). Further, the Gaussianised data in the merged cluster (2, 3, 4) can be optimally fit with five factors, explaining about $93.3\%$ of the total variance. Table~\ref{tab:grb-comp-loadings} presents the estimated factor loadings after quartimax rotation~\citep{costello2005}, where the first factor loading vector for the composite cluster (2, 3, 4) is a weighted average in Gaussianised space with low to high contributions from $T_{90}$, $F_4$, $F_1$, $F_2$, $F_3$, $P_{64}$, $P_{256}$ and $P_{1024}$ in that order. The second factor is mainly determined by the two flux arrival times along with small contributions from the four time-integrated fluences in Gaussianised space, and minor amounts of the three peak fluxes from the opposite direction. The third factor loading vector is a contrast between a weighted average of the first three time-integrated fluences (with $F_3$ contributing around $60\%$ less than $F_1$ and $F_2$) on the one hand, and a simple average of minor contributions from the three peak fluxes on the other, all in Gaussianised space. The fourth factor is again a contrast between a weighted average of small contributions from $F_3$ and $F_4$ in Gaussianised space, and a simple average of minor contributions from $P_{64}$, $P_{256}$ and $P_{1024}$. Finally, the fifth factor has small but non-negligible contributions (in opposing sign) from $P_{64}$ and $P_{1024}$. In addition, the averaged factor scores estimated from the composite cluster (2, 3, 4) are given in Table~\ref{tab:grb-5g-comp-fascore}, where the magnitudes of the score values indicate that, for GRBs from this compound cluster, the importance ratings on the second and third factors are close to the overall average rating level while the the first and last factors have notably higher rating scores, specifically, factor 1 $>$ factor 5 $>$ factor 4 $>$ factor 2 $>$ factor 3. We can see that the composite cluster (2, 3, 4) present distinct characteristics from clusters 1 and 5. 

Having explained the composite cluster (2, 3, 4), we now move towards explaining the composite cluster (1, 5) formed after the second stage of merging. We performed factor analysis on the GRBs from the composite cluster (1, 5) in Gaussianised space, and obtained an optimal four-factor model that explains $91.9\%$ of the total variances. The group-wise estimated factor loadings under quartimax rotation are given in Table~\ref{tab:grb-comp-loadings} with non-negligible values, where the first factor is seen to have small contributions from the two flux arrival times, moderate contributions from $F_4$, substantial contributions from $F_1$, $F_2$, $F_3$ and $P_{64}$, and large contributions from $P_{256}$ and $P_{1024}$ (all in Gaussianised space). The second factor loading is majorly dominated by the two flux arrival times, along with moderate to small contributions from the four time-integrated fluences and $P_{1024}$, and moderate to marginal contributions from $P_{64}$ and $P_{256}$ in the opposite direction. The third factor loading vector in Gaussianised space is a small contrast among the last three time-integrated fluences and $P_{64}$, with a weighted average of $F_3$ and $F_4$ on one hand where the latter contributes almost twice in magnitude to that of the former, and small contributions from $F_2$ and $P_{64}$ on the other side. The last and fourth factor is a contrast (again in Gaussianised space) between small amounts of $P_{64}$ and $P_{256}$ and a simple average of small contributions from $F_2$ and $F_3$. Table~\ref{tab:grb-5g-comp-fascore} again provides the averaged factor scores for the compound cluster (1, 5), where the importance of the four latent factor variables are rated as factor 1 $>$ factor 3 $>$ factor 4 $>$ factor 2, with the first factor's rating $50\%$ higher than the rating on the last factor.

\begin{table}
\centering
\caption{The mean factor scores for the two composite clusters (1, 5) and (2, 3, 4).}
\label{tab:grb-5g-comp-fascore} 
\addtolength{\tabcolsep}{-0.6em}
\begin{tabular}{c|c|c|c|c|c}
\hline\hline
\bf{Cluster}
&\bf{Factor 1}&\bf{Factor 2}& \bf{Factor 3}& \bf{Factor 4}& \bf{Factor 5}\\
\hline\hline
{({\bf\textcolor{err1}{1}},{\bf\textcolor{err6}{5}})} 
&0.167 &0.111 &0.118 &0.112 &--\\
{({\bf\textcolor{err2}{2}}, {\bf\textcolor{err3}{3}}, {\bf\textcolor{err5}{4}})}  
&0.028  &0.006 &-0.003  &0.019  &0.025\\
\hline
\end{tabular}
\end{table}
\subsection{Additional interpretation of our results}
\label{sec:3parscharacterisation}
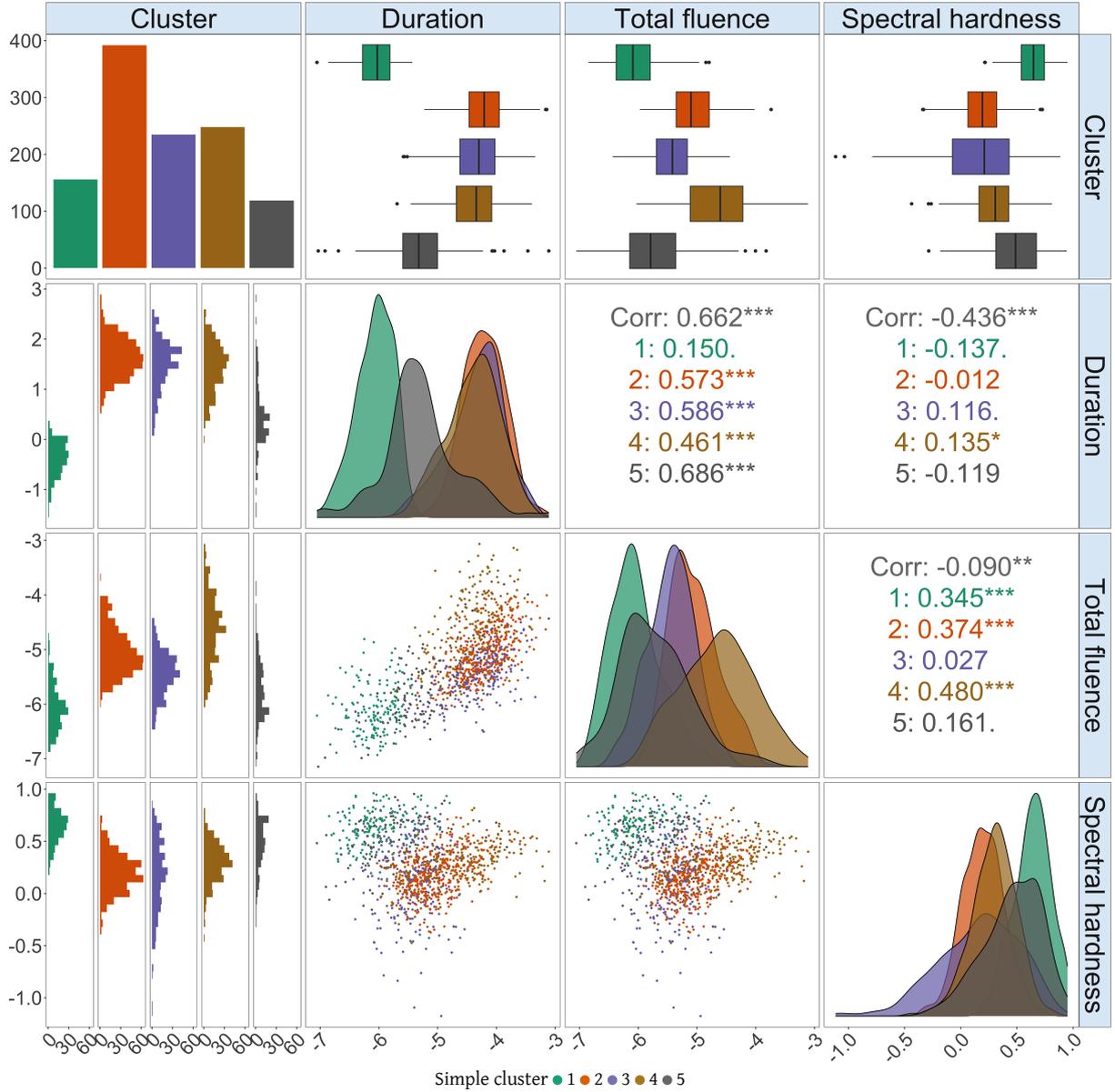
\begin{figure*}
    \centering
   \includegraphics[width = 0.9\textwidth]{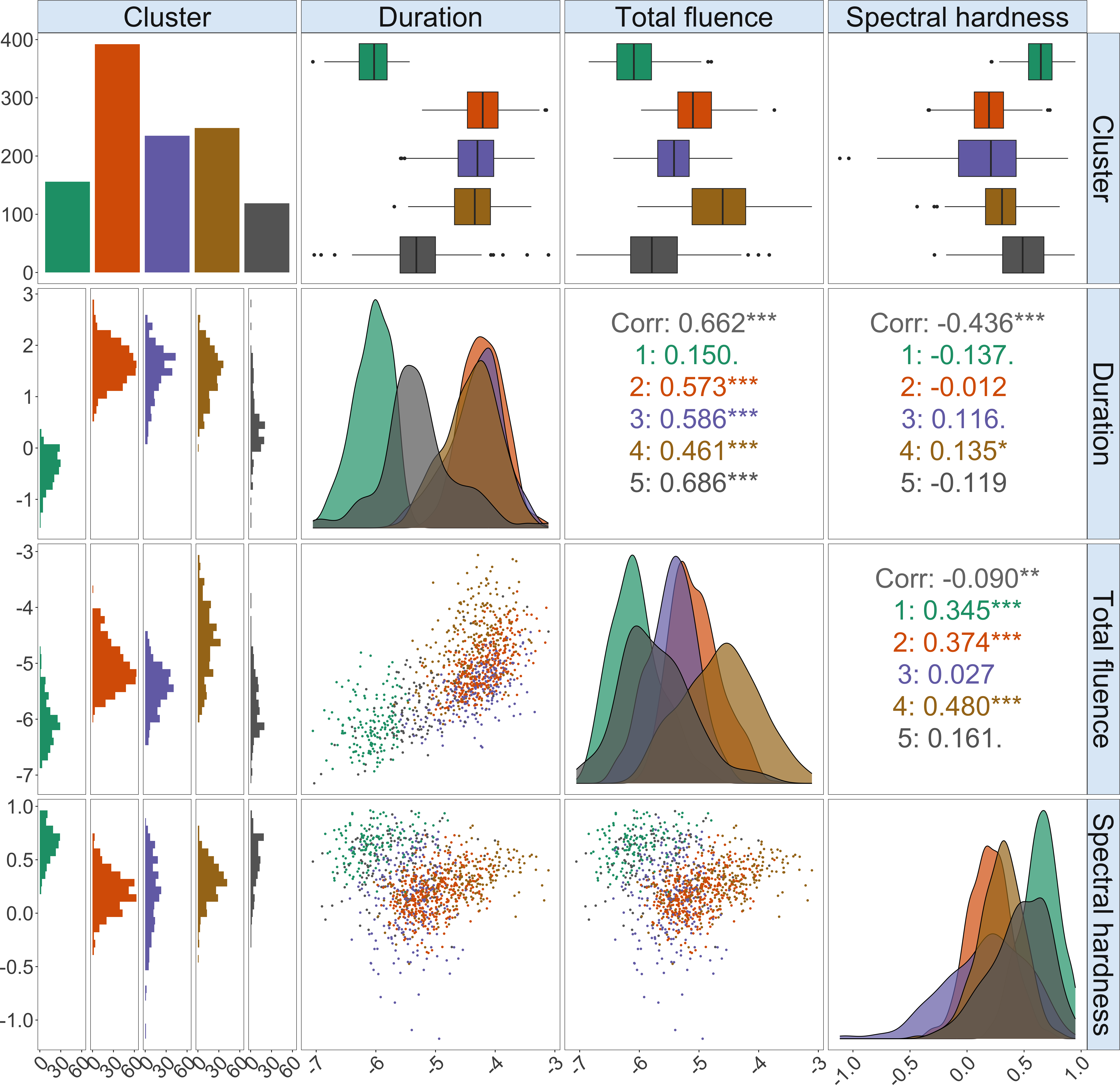}\\ 
    \centering
    \text{
     Simple cluster \tikz\draw[err1,fill=err1] (0,0) circle (.5ex); 1
     \tikz\draw[err2,fill=err2] (0,0) circle (.5ex); 2
     \tikz\draw[err3,fill=err3] (0,0) circle (.5ex); 3
     \tikz\draw[err5,fill=err5] (0,0) circle (.5ex); 4
     \tikz\draw[err6,fill=err6] (0,0) circle (.5ex); 5
    }
    \caption{Densities and scatter plots of the three derived features (in $\log_{10}$ scale): duration, total fluence and spectral hardness for the 1150 thinned complete BATSE Catalog GRBs with colors indicating the five simple clusters. Correlations between features are shown in the upper panel. Note the lack of skewness in the duration variable in the first cluster, and the very high skewness in the last cluster, which distinguishes the two groups.}
    \label{fig:grb-5g-features-dist}
\end{figure*}
\citet{mukherjeeetal98} provided a Duration-Fluence-Spectrum scheme for characterising the BATSE GRBs using one original and two composite features. These three summary features are duration ($T_{90}$), total fluence ($F_1+F_2+F_3+F_4$) and spectral hardness ($H_{321} = F_3/(F_1+F_2)$). This classification and characterisation scheme was also adopted by \citet{chattopadhyayetal07} and in several subsequent studies~\citep{chattopadhyayandmaitra17,chattopadhyayandmaitra18,berryandmaitra19,almodovarandmaitra20,gorenandmaitra22}. We display the distribution of duration, total fluence and spectral hardness in each simple group by means of density and pairwise scatter plots in Figure~\ref{fig:grb-5g-features-dist}. While pairwise and individual density plots provide only a partial understanding of all the relationships, we can see very well both the distinctiveness of the five simple groups, and also similarities between some of them that lead to their merging at a higher level, which we now explain.

For clusters 2, 3, 4 and 5, duration has a stronger correlation with total fluence compared to cluster 1 where total fluence is weakly correlated with duration while negatively related to spectral hardness. The correlation between duration and spectral hardness is negligible for cluster 2. Comparing the distributions of the three features across the five clusters, we see that, for clusters 2, 3 and 4, the duration and total fluence are at higher levels than clusters 1 and 5 with the duration for cluster 5 presenting the largest variation, while the spectral hardness levels are lower for clusters 2, 3 and 4 compared to clusters 1 and 5, which gives insights on understanding the composite cluster (2, 3, 4) and two simple clusters 1 and 5 obtained after the first stage of merging in Section~\ref{sec:grb-mobsync}. Our observations lead to the categorization of the five simple clusters in Table~\ref{tab:grb-5g-features}, where the means indicate clusters 1 and 5 to be shorter-faint-hard and short-faint-hard, while clusters 2 (long-intermediate-soft), 3 (long-intermediate-intermediate) and 4 (long-bright-intermediate) show similar properties that are more distinct from clusters 1 and 5, which is indicated earlier by the final merging results in Section~\ref{sec:grb-mobsync}. At the same time, it should be noted that the skewness and variation in the duration parameter for the two clusters 1 and 5 are markedly different, with the first cluster having virtually less skew, but the fifth cluster more skewed and highly variable, which also provides understanding for why these two groups are separately identified as simple (ellipsoidal) clusters. 
\begin{table}
\centering
\caption{\label{tab:grb-5g-features} Characterization of the five simple GRB clusters by the Duration-Fluence-Spectrum scheme.}
\addtolength{\tabcolsep}{-0.4em}
\begin{tabular}{c|c|c|c|c|c}
\hline\hline
\bf{Cluster} & 
\bf{\textcolor{err1}{1}} & \bf{\textcolor{err2}{2}} & \bf{\textcolor{err3}{3}} & \bf{\textcolor{err5}{4}} & \bf{\textcolor{err6}{5}}\\ 
    \hline\hline
\bf{Duration} & shorter &long &long &long &short \\
\hline
\bf{Fluence} & faint &intermediate &intermediate
&bright  &faint \\
\hline
\bf{Spectrum} & hard 
&soft
&intermediate 
&intermediate &hard \\
\hline
\end{tabular}
\end{table}

\section{Conclusions}
\label{sec:con}
Traditional classification and characterisation of GRBs, largely using the duration parameter, have typically been seen to find  two or three classes of GRBs, while recent studies~\citep{chattopadhyayandmaitra17,chattopadhyayandmaitra18,berryandmaitra19,almodovarandmaitra20}
have identified and characterised new GRB categories by including more parameters and using more advanced statistical techniques. In light of these conflicting results, we investigated 1150 GRBs in the complete BATSE Catalog with all nine astrophysical parameters $T_{50}, T_{90}, F_1, F_2, F_3, F_4, P_{64}, P_{256}, P_{1024}$ that are greater than 2$\sigma$ for the flux arrival times, and over 1$\sigma$ for the time-integrated fluences and the peak flux parameters. Our analysis classified these 1150 bursts into five ellipsoidal-shaped groups using MBC with a Gaussian mixture of factor analysers, where the model was applied with matrix-free computations that allow efficient parameter estimation. Following the Duration-Fluence-Spectrum scheme~\citep{mukherjeeetal98,chattopadhyayetal07}, the five GRB groups were further distinguished as shorter-faint-hard with 156 GRBs from group 1 and 119 short-faint-hard GRBs from group 5 with the fifth group having a longer skew and more variation in duration, long-intermediate-soft with 392 GRBs from group 2, long-intermediate-intermediate with 235 GRBs from group 3, and long-bright-intermediate with 248 GRBs from group 4, in terms of the duration ($T_{90}$), total fluence
($F_1+F_2+F_3+F_4$) and spectral hardness ratio ($H_{321}=F_3/(F_1+F_2)$). These groups were also distinctly characterised in terms of five latent factors of the nine variables. Our findings here provide support to these recent findings that there are five ellipsoidal groups in the BATSE GRB dataset. 
Qualitatively similar results were obtained with the unthinned sample of 1598 GRBs from the complete BATSE  catalogue that ignored measurement error while including GRBs for analysis. Once again, we obtained five simple groups which merged first to three and then two compound groups in a similar manner as with this thinned sample that we  studied in detail in this paper.

However, these five ellipsoidal groups have some discordance with other studies that find a fewer number of groups of GRBs when using a few parameters in the analysis. To understand the discrepancy, and motivated by the syncytial clustering methods of \cite{almodovarandmaitra20,chattopadhyayandmaitra18,chattopadhyayetal22}, we developed MOBSynC that merges less-separated groups into composite clusters based on the overlap measures between individual groups. When applied to the five simple groups found by MixFAD, we saw that MOBSynC first formed one larger cluster of the second, third and forth simple groups, and left the other two simple groups unchanged. This yielded a three-group characterisation of the GRBs after the first stage of merging, that can mostly be characterised in terms of duration and in terms of long, short and very short durations ($T_{90}$). The next stage encourages the two simple groups to merge, yielding a two-groups characterisation of the GRBs.
Our results show a multi-layered characterisation of the BATSE GRBs, that is also supported by the characterisation in terms of the  Duration-Fluence-Spectrum scheme. Our analysis therefore shows that the presence of two composite general-shaped kinds of GRBs that are each themselves composed of two and three ellipsoidally-shaped sub-groups. The two simple groups are more separable than the three simple groups, and form a three-groups solution at an intermediate stage. At the deepest level, we get a characterisation in terms of five ellipsoidal groups. Thus, we explain the findings of two, three or five kinds of GRBs in the literature, and show that all three solutions are explained in a syncytial setup.

Besides identifying the different kinds of GRBs, we also obtained group-wise characterisation of the variability in the simple groups of GRBs as a natural by-product of our MixFAD algorithm, and found that the variability in the nine parameters can be explained in terms of a few latent factors. The identified GRB groups show distinguished dependence structures among the parameters via the underlying features, and different importance ratings on the factors. At the higher level, the latent factor representation of the compound groups is not immediate, so we developed a novel approach that Gaussianised the data in each composite cluster and explained the variability in these compound groups in terms of four and five latent factors. 

There are several potential topics that arise as worthy of investigation and extension beyond the current work. First, as pointed out by the reviewer, the parameters in each  GRB are observed with measurement (or systematic) error. While these errors are available and should be incorporated in our analysis, we were not able to incorporate them in our analysis because there currently exists no statistical methodology that allows for their inclusion in a MixFAD clustering framework that would also allow us not just to group the GRBs, but also to explain their variability in terms of a few underlying group-specific latent factors. Such statistical methodology is very important to develop, validate and implement on the GRB dataset, and is left for future research. Secondly, 
the BATSE Catalog also contains GRBs observed with partially recorded features, and clustering approaches should take the presence of missing values into consideration. However, while methods to cluster such data using general mixture models~\citep{gorenandmaitra22} exist, methodology to cluster such data using a mixture of factor analysers needs development and implementation. It would be interesting to see if including these partially observed GRBs in the analysis can provide additional insight into the dataset. Further, our current model assumes same number of factors for all clusters, but it may be worth considering a more flexible model that allows for varying the  number of factors across groups. Finally, this paper has provided an approach to characterising and clustering GRBs: we believe that our method is general enough to also apply to other similar datasets to reveal complicated structures and latent relationships. 

\section*{Acknowledgments}
The authors are very grateful to an anonymous reviewer whose speedy but careful review and insightful comments on two earlier versions of this manuscript greatly improved its content. 
Our thanks also to the Scientific Editor and the Assistant Editor for their timely help in the review process.




\section*{Data Availability Statement}
The simulated dataset used in Section~\ref{sec:illustration} of this article are publicly available at \url{https://github.com/fanstats/MixFAD-GRB}. The GRB dataset used in this article contains the complete data on all nine parameters that was obtained in raw form from the complete BATSE GRB catalogue at 
\url{https://heasarc.gsfc.nasa.gov/cgi-bin/W3Browse/w3query.pl?&tablehead=name%3Dheasarc_batsegrb%26description%3DCGRO%2FBATSE+Gamma-Ray+Burst+Catalog%26url%3Dhttp%3A%2F%2Fheasarc.gsfc.nasa.gov%2FW3Browse%2Fcgro%2Fbatsegrb.html%26archive%3DY%26radius%3D300%26mission%3DCGRO%26priority%3D3&mission=CGRO&Action=More+Options&Action=Parameter+Search&ConeAdd=1}. Both the raw and processed GRB datasets are also available at \url{https://github.com/fanstats/MixFAD-GRB}. 
	Software for performing {\tt MixFAD} will be made publicly available as a \texttt{R}~\citep{R} package under the same name, while {\tt MOBSynC} will be available as part of the publicly available {\tt SynClustR} package in {\tt R}~\citep{R}.

\bibliographystyle{mnras}
\bibliography{references} 

@article{zhuetal24,
author = {Zhu, Si-Yuan and Sun, Wan-Peng and Ma, Da-Ling and Zhang, Fu-Wen},
year = {2024},
month = {06},
pages = {1434-1443},
title = {Classification of Fermi gamma-ray bursts based on machine learning},
volume = {532},
journal = {Monthly Notices of the Royal Astronomical Society},
doi = {10.1093/mnras/stae1594}
}

@article{steinhardtetal23,
author = {Steinhardt, Charles and Mann, William and Rusakov, Vadim and Jespersen, Christian},
year = {2023},
month = {03},
pages = {67},
volume = {945},
journal = {The Astrophysical Journal},
title = {Classification of BATSE, Swift, and Fermi Gamma-Ray Bursts from Prompt Emission Alone},
doi = { 10.3847/1538-4357/acb999}
}

@article{yangetal22,
author = {Yang, Haifeng and Shi, Chenhui and Cai, Jianghui and Zhou, Lichan and Yang, Yuqing and Zhao, Xujun and He, Yanting and Hao, Jing},
year = {2022},
month = {09},
pages = {5496-5523},
title = {Data mining techniques on astronomical spectra data – I. Clustering analysis},
volume = {517},
journal = {Monthly Notices of the Royal Astronomical Society},
doi = {10.1093/mnras/stac2975}
}

@article{tothetal19,
    author = {T\'oth, B G and R\'acz, I I and Horv\'ath, I},
    title = "{Gaussian-mixture-model-based cluster analysis of gamma-ray bursts in the BATSE catalog}",
    journal = {Monthly Notices of the Royal Astronomical Society},
    volume = {486},
    number = {4},
    pages = {4823-4828},
    year = {2019},
    month = {05},
    issn = {0035-8711},
    doi = {10.1093/mnras/stz1188},
    url = {https://doi.org/10.1093/mnras/stz1188},
    eprint = {https://academic.oup.com/mnras/article-pdf/486/4/4823/28599153/stz1188.pdf},
}

@article{petersonetal18,
        author = {A. D. Peterson and A. P. Ghosh and R. Maitra},
        title = {Merging $K$-means with hierarchical clustering for identifying general-shaped groups},                         
        doi = {10.1002/sta4.172},
        journal = {Stat},
        year = {2018},                             
        volume = {7},                   
        number = {1},
        pages = {e172},
}

@book{nelsen06,                                                                           
author = {Nelsen, R. B.},                                                                 
year = {2006},                                    
title = {An Introduction to Copulas},
edition = {2},                                                                                 
publisher = {Springer},                                                         
address = {New York}
}

@article{yuanetal18,
doi = {10.3847/1538-4365/aaed3b},
url = {https://dx.doi.org/10.3847/1538-4365/aaed3b},
year = {2018},
month = {dec},
publisher = {The American Astronomical Society},
volume = {239},
number = {2},
pages = {33},
author = {Zunli Yuan and Jiancheng Wang and D. M. Worrall and Bin-Bin Zhang and Jirong Mao},
title = {Determining the Core Radio Luminosity Function of Radio AGNs via Copula},
journal = {The Astrophysical Journal Supplement Series}
}

@ARTICLE{hennig10,
  author = {Hennig, C.},
  title = {Methods for merging {G}aussian mixture components},
  journal = {Advances in Data Analysis and Classification},
  year = {2010},
  doi = {10.1007/s11634-010-0058-3}
}

@article{thurstone31,
	author = {Thurstone, Louis},
	year = {1931},
	title = {Multiple factor analysis},
	journal = {Psychological Review},
	volume = {38},
	number = {5},
	pages = {406-427},
	doi = {10.1037/h0069792}
}

@article{berryandmaitra19,
	author = {N. Berry and R. Maitra},
	title = {{TiK}-means: Transformation-infused $K$-means clustering for skewed groups},
	year = {2019},
	volume = {12},
	number = {3},
	pages = {223-233},
	journal = {Statistical Analysis and Data Mining -- The ASA Data Science Journal},
	doi = {10.1002/sam.11416},
}

@article{tarnopolski19,
doi = {10.3847/1538-4357/ab4fe6},
url = {https://dx.doi.org/10.3847/1538-4357/ab4fe6},
year = {2019},
month = {dec},
publisher = {The American Astronomical Society},
volume = {887},
number = {1},
pages = {97},
author = {Mariusz Tarnopolski},
title = {Multivariate Analysis of BATSE Gamma-Ray Burst Properties Using Skewed Distributions},
journal = {The Astrophysical Journal},
}

@article{tarnopolski22,
	author = {Tarnopolski, Mariusz},
	title = {Graph-based clustering of gamma-ray bursts},
	DOI= "10.1051/0004-6361/202038645",
	url= "https://doi.org/10.1051/0004-6361/202038645",
	journal = {A&A},
	year = 2022,
	volume = 657,
	pages = "A13",
}

@article{distefanoetal09,
  title={Understanding and Using Factor Scores: Considerations for the Applied Researcher},
  author={Christine Distefano and Min Zhu and Diana Mindrila},
  journal={Practical Assessment, Research and Evaluation},
  year={2009},
  volume={14},
  number = {1},  
  pages={20}
}

@article{bhaveetal22,
author = {Bhave, A. and Kulkarni, S. and Desai, S. and Srijith, P. K.},
title = {Two dimensional clustering of Gamma-Ray Bursts using durations and hardness},
journal = {Astrophysics and Space Science},
volume = {367},
number = {39},
year = {2022},
doi = {https://doi.org/10.1007/s10509-022-04068-z}
}

@article{bhardwajetal23,
    author = {Bhardwaj, Shubham and Dainotti, Maria G and Venkatesh, Sachin and Narendra, Aditya and Kalsi, Anish and Rinaldi, Enrico and Pollo, Agnieszka},
    title = "{GRB optical and X-ray plateau properties classifier using unsupervised machine learning}",
    journal = {Monthly Notices of the Royal Astronomical Society},
    volume = {525},
    number = {4},
    pages = {5204-5223},
    year = {2023},
    month = {08},
    issn = {0035-8711},
    doi = {10.1093/mnras/stad2593},
    url = {https://doi.org/10.1093/mnras/stad2593},
    eprint = {https://academic.oup.com/mnras/article-pdf/525/4/5204/51542203/stad2593.pdf},
}

@ARTICLE{mazetsetal81,
   author = {{Mazets}, E.~P. and {Golenetskii}, S.~V. and {Ilinskii}, V.~N. and 
	{Panov}, V.~N. and {Aptekar}, R.~L. and {Gurian}, I.~A. and 
	{Proskura}, M.~P. and {Sokolov}, I.~A. and {Sokolova}, Z.~I. and 
	{Kharitonova}, T.~V.},
    title = "{Catalog of cosmic gamma-ray bursts from the KONUS experiment data. I.}",
  journal = {\apss},
     year = 1981,
    month = nov,
   volume = 80,
    pages = {3-83},
      doi = {10.1007/BF00649140},
   adsurl = {http://adsabs.harvard.edu/abs/1981Ap%26SS..80....3M}
}

@INPROCEEDINGS{dezalayetal92,
   author = {{Dezalay}, J.-P. and {Barat}, C. and {Talon}, R. and {Syunyaev}, R. and 
	{Terekhov}, O. and {Kuznetsov}, A.},
    title = "{Short cosmic events - A subset of classical GRBs?}",
booktitle = {American Institute of Physics Conference Series},
     year = 1992,
   series = {American Institute of Physics Conference Series},
   volume = 265,
   editor = {{Paciesas}, W.~S. and {Fishman}, G.~J.},
    pages = {304-309},
   adsurl = {http://adsabs.harvard.edu/abs/1992AIPC..265..304D}
}

@ARTICLE{norrisetal84,
   author = {{Norris}, J.~P. and {Cline}, T.~L. and {Desai}, U.~D. and {Teegarden}, B.~J.
	},
    title = "{Frequency of fast, narrow gamma-ray bursts}",
  journal = {\nat},
     year = 1984,
    month = mar,
   volume = 308,
    pages = {434},
      doi = {10.1038/308434a0},
   adsurl = {http://adsabs.harvard.edu/abs/1984Natur.308..434N}
}

@article{horvathetal04,
   author = {{Horv{\'a}th}, I. and {M{\'e}sz{\'a}ros}, A. and {Bal{\'a}zs}, L.~G. and 
	{Bagoly}, Z.},
    title = "{Where is the Third Subgroup of Gamma-Ray Bursts?}",
  journal = {Baltic Astronomy},
   eprint = {astro-ph/0507688},
     year = 2004,
   volume = 13,
    pages = {217-220},
   adsurl = {http://adsabs.harvard.edu/abs/2004BaltA..13..217H}
}

@ARTICLE{paczynski98,
   author = {{Paczy{\'n}ski}, B.},
    title = "{Are Gamma-Ray Bursts in Star-Forming Regions?}",
  journal = {\apjl},
   eprint = {astro-ph/9710086},
     year = 1998,
    month = feb,
   volume = 494,
    pages = {L45-L48},
      doi = {10.1086/311148},
   adsurl = {http://adsabs.harvard.edu/abs/1998ApJ...494L..45P},
}

@article{baudryetal10,
author = {Jean-Patrick Baudry and Adrian E. Raftery and Gilles Celeux and Kenneth Lo and RaphaÃ«l Gottardo},
title = {Combining Mixture Components for Clustering},
journal = {Journal of Computational and Graphical Statistics},
volume = {19},
number = {2},
pages = {332-353},
year = {2010},
doi = {10.1198/jcgs.2010.08111},

URL = { 
        http://dx.doi.org/10.1198/jcgs.2010.08111
    
},
eprint = { 
        http://dx.doi.org/10.1198/jcgs.2010.08111
    
}
}

@Article{horvathandtoth16,
author="Horv{\'a}th, I.
and T{\'o}th, B. G.",
title="The duration distribution of Swift Gamma-Ray Bursts",
journal={\apss},
year="2016",
volume="361",
number="5",
pages="155",
issn="1572-946X",
doi="10.1007/s10509-016-2748-6",
url="http://dx.doi.org/10.1007/s10509-016-2748-6"
}

@Article{melnykovandmaitra11,
  Title                    = {{CARP}: Software for Fishing Out Good Clustering Algorithms},
  Author                   = {Melnykov, V. and Maitra, R.},
  Journal                  = {Journal of Machine Learning Research},
  Year                     = {2011},
  Pages                    = {69 - 73},
  Volume                   = {12}
}

@Book{mclachlanandpeel00,
  Title                    = {Finite Mixture Models},
  Author                   = {G. McLachlan and D. Peel},
  Publisher                = {John Wiley and Sons, Inc.},
  Year                     = {2000},

  Address                  = {New York},
  doi={10.1002/0471721182}
}

@Book{mclachlanandkrishnan08,
  Title                    = {The {EM} Algorithm and Extensions},
  Author                   = {McLachlan, G. and Krishnan, T. },
  Publisher                = {Wiley},
  Year                     = {2008},

  Address                  = {New York},
  Edition                  = {Second},
  doi={10.2307/2534032}
}

@ARTICLE{dempsteretal77,
  author = {Dempster, A. P. and Laird, N. M. and Rubin, D. B.},
  title = {Maximum Likelihood for Incomplete Data via the {EM} Algorithm (with
	discussion)},
  journal = {Jounal of the Royal Statistical Society, Series B},
  year = {1977},
  volume = {39},
  pages = {1-38},
  url={http://www.jstor.org/stable/2984875},
  doi={10.2307/2984875}
}

@article{melnykov16,
author = {Melnykov, Volodymyr},
year = {2016},
month = {05},
pages = {66-90},
title = {Merging Mixture Components for Clustering Through Pairwise Overlap},
volume = {25},
number = {1},
journal = {Journal of Computational and Graphical Statistics},
doi = {10.1080/10618600.2014.978007}
}

@article{gorenandmaitra22,    
        author = {Goren, E. M. and Maitra, R.},
        title = {Fast Model-based Clustering of Partial Records},
        year = {2022},
        journal = {Stat},   
        volume = {11},                               
        number = {1},                          
        pages = {416},
        doi = {10.1002/sta4.416}
        }

@Manual{R,
    title = {R: A Language and Environment for Statistical Computing},
    author = {{R Core Team}},
    organization = {R Foundation for Statistical Computing},
    address = {Vienna, Austria},
    year = {2024},
    url = {https://www.R-project.org/},
  }

@article{bagolyetal98,
  author={Z. Bagoly and A. M\'esz\'aros and I. Horv\'ath and L. G. Bal\'azs and P. M\'esz\'aros},
  title={A Principal Component Analysis of the 3B Gamma-Ray Burst Data},
  journal={\apj},
  volume={498},
  number={1},
  pages={342},
  url={http://stacks.iop.org/0004-637X/498/i=1/a=342},
  year={1998},
  doi={	10.1086/305530}
}

@article{bagolyetal09,
  title={Factor analysis of the long gamma-ray bursts},
  author={Z. Bagoly and Luis Borgonovo and A. M{\'e}sz{\'a}ros and Lajos G. Bal{\'a}zs and Istv{\'a}n Horvath},
  journal={Astronomy and Astrophysics},
  year={2009},
  volume={493},
  pages={51-54}, 
  url={https://api.semanticscholar.org/CorpusID:16295856}
}

@article{chattopadhyayetal07,
  author={Tanuka Chattopadhyay and Ranjeev Misra and Asis Kumar Chattopadhyay and Malay Naskar},
  title={Statistical Evidence for Three Classes of Gamma-Ray Bursts},
  journal={\apj},
  volume={667},
  number={2},
  pages={1017},
  url={http://stacks.iop.org/0004-637X/667/i=2/a=1017},
 doi = {https://doi.org/10.1086/520317},
year={2007}
}

@article{hakkilaetal03,
  author={Jon Hakkila and Timothy W. Giblin and Richard J. Roiger and David J. Haglin and William S. Paciesas and Charles A.
Meegan},
  title={How Sample Completeness Affects Gamma-Ray Burst Classification},
  journal={\apj},
  volume={582},
  number={1},
  pages={320},
  url={http://stacks.iop.org/0004-637X/582/i=1/a=320},
  year={2003},
  doi={10.1086/344568}
}

@article{horvathetal98,
  author={I. Horv\'ath},
  title={A Third Class of Gamma-Ray Bursts?},
  journal={\apj},
  volume={508},
  number={2},
  pages={757},
  url={http://stacks.iop.org/0004-637X/508/i=2/a=757},
  year={1998},
  doi={10.1086/306416}
}

@ARTICLE{kouveliotouetal93,
   author = {{Kouveliotou}, C. and {Meegan}, C.~A. and {Fishman}, G.~J. and 
	{Bhat}, N.~P. and {Briggs}, M.~S. and {Koshut}, T.~M. and {Paciesas}, W.~S. and 
	{Pendleton}, G.~N.},
    title = "{Identification of two classes of gamma-ray bursts}",
  journal = {\apjl},
     year = 1993,
    month = aug,
   volume = 413,
    pages = {L101-L104},
      doi = {10.1086/186969},
   adsurl = {http://adsabs.harvard.edu/abs/1993ApJ...413L.101K}
}

@article{piran05,
  title = {The physics of gamma-ray bursts},
  author = {Piran, Tsvi},
  journal = {Rev. Mod. Phys.},
  volume = {76},
  issue = {4},
  pages = {1143--1210},
  numpages = {0},
  year = {2005},
  month = {Jan},
  publisher = {American Physical Society},
  doi = {10.1103/RevModPhys.76.1143},
  url = {http://link.aps.org/doi/10.1103/RevModPhys.76.1143}
}

@ARTICLE{piran92,
   author = {{Piran}, T.},
    title = "{The implications of the Compton (GRO) observations for cosmological gamma-ray bursts}",
  journal = {apjl},
     year = 1992,
    month = apr,
   volume = 389,
    pages = {L45-L48},
      doi = {10.1086/186345},
   adsurl = {http://adsabs.harvard.edu/abs/1992ApJ...389L..45P}
}

@article{woosleyandbloom06,
author = {S.E. Woosley and J.S. Bloom},
title = {The Supernova–Gamma-Ray Burst Connection},
journal = {\araa},
volume = {44},
number = {1},
pages = {507-556},
year = {2006},
doi = {10.1146/annurev.astro.43.072103.150558},

URL = { http://dx.doi.org/10.1146/annurev.astro.43.072103.150558},
eprint = { http://dx.doi.org/10.1146/annurev.astro.43.072103.150558}

}

@Article{zitounietal15,
author="Zitouni, H.
and Guessoum, N.
and Azzam, W. J.
and Mochkovitch, R.",
title="Statistical study of observed and intrinsic durations among BATSE and Swift/BAT GRBs",
journal={\apss},
year="2015",
volume="357",
number="1",
pages="7",
issn="1572-946X",
doi="10.1007/s10509-015-2311-x",
url="http://dx.doi.org/10.1007/s10509-015-2311-x"
}

@article{nakar07,
title = "Short-hard gamma-ray bursts ",
journal = "Physics Reports ",
volume = "442",
number = "1–6",
pages = "166 - 236",
year = "2007",
note = "The Hans Bethe Centennial Volume 1906-2006 ",
issn = "0370-1573",
doi = "http://dx.doi.org/10.1016/j.physrep.2007.02.005",
url = "http://www.sciencedirect.com/science/article/pii/S0370157307000476",
author = "Ehud Nakar"
}

@article{ horvath02,
	author = {Horv\'ath, I.},
	title = {A further study of the BATSE Gamma-Ray Burst duration 
distribution},
	DOI= "10.1051/0004-6361:20020808",
	url= "http://dx.doi.org/10.1051/0004-6361:20020808",
	journal = {\aap},
	year = 2002,
	volume = 392,
	number = 3,
	pages = "791-793",
	month = "",
}

@ARTICLE {horvath09,
    author  = "Horv\'ath, I.",
    title   = "Classification of BeppoSAX's gamma-ray bursts",
    journal = {\apss},
    year    = "2009",
    volume  = "323",
    number  = "83-86",
    doi={10.1007/s10509-009-0039-1}
}

@article{ tarnopolski15,
	author = {Tarnopolski, M.},
	title = {Analysis of Fermi gamma-ray burst duration distribution},
	DOI= "10.1051/0004-6361/201526415",
	url= "http://dx.doi.org/10.1051/0004-6361/201526415",
	journal = {\aap},
	year = 2015,
	volume = 581,
	pages = "A29",
	month = "",
}

@ARTICLE{mukherjeeetal98,
   author = {{Mukherjee}, S. and {Feigelson}, E.~D. and {Jogesh Babu}, G. and 
	{Murtagh}, F. and {Fraley}, C. and {Raftery}, A.},
    title = "{Three Types of Gamma-Ray Bursts}",
  journal = {\apj},
   eprint = {astro-ph/9802085},
     year = 1998,
    month = nov,
   volume = 508,
    pages = {314-327},
      doi = {10.1086/306386},
   adsurl = {http://adsabs.harvard.edu/abs/1998ApJ...508..314M}
}

@ARTICLE{maitra09,
author={R. Maitra},
journal={IEEE/ACM Transactions on Computational Biology and Bioinformatics},
title={Initializing Partition-Optimization Algorithms},
year={2009},
volume={6},
number={1},
pages={144-157},
doi={10.1109/TCBB.2007.70244},
ISSN={1545-5963},
month={Jan},}

@article{melnykovandmaitra10,
author = "Melnykov, Volodymyr and Maitra, Ranjan",
doi = "10.1214/09-SS053",
fjournal = "Statistics Surveys",
journal = "Statist. Surv.",
pages = "80--116",
publisher = "The American Statistical Association, the Bernoulli Society, the Institute of Mathematical Statistics, and the Statistical Society of Canada",
title = "Finite mixture models and model-based clustering",
url = "http://dx.doi.org/10.1214/09-SS053",
volume = "4",
year = "2010"
}

@article{almodovarandmaitra20,
  author  = {Israel A. Almodovar-Rivera and Ranjan Maitra},
  title   = {Kernel-estimated Nonparametric Overlap-Based Syncytial Clustering},
  journal = {Journal of Machine Learning Research},
  year    = {2020},
  volume  = {21},
  number  = {122},
  pages   = {1--54},
  url     = {http://jmlr.org/papers/v21/18-435.html}
}

@article{maitraandmelnykov10,
author = {Ranjan Maitra and Volodymyr Melnykov},
title = {Simulating Data to Study Performance of Finite Mixture Modeling and Clustering Algorithms},
journal = {Journal of Computational and Graphical Statistics},
volume = {19},
number = {2},
pages = {354-376},
year = {2010},
doi = {10.1198/jcgs.2009.08054},

URL = { 
        http://dx.doi.org/10.1198/jcgs.2009.08054
    
},
eprint = { 
        http://dx.doi.org/10.1198/jcgs.2009.08054
    
}
}

@article{maitra10,
title = "A re-defined and generalized percent-overlap-of-activation measure for studies of fMRI reproducibility and its use in identifying outlier activation maps ",
journal = "NeuroImage ",
volume = "50",
number = "1",
pages = "124 - 135",
year = "2010",
note = "",
issn = "1053-8119",
doi = "http://dx.doi.org/10.1016/j.neuroimage.2009.11.070",
url = "http://www.sciencedirect.com/science/article/pii/S1053811909012567",
author = "Ranjan Maitra"
}

@Article{melnykovetal12,
    author = {Volodymyr Melnykov and Wei-Chen Chen and Ranjan Maitra},
   title = {{MixSim}: An R Package for Simulating Data to Study Performance of Clustering Algorithms},
   journal = {Journal of Statistical Software},
   volume = {51},
   number = {1},
   year = {2012},
   issn = {1548-7660},
   pages = {1--25},
   doi = {10.18637/jss.v051.i12},
   url = {https://www.jstatsoft.org/index.php/jss/article/view/v051i12}
}

@article{pendletonetal97,
 author={G. N. Pendleton and W. S. Paciesas and M. S. Briggs and R. D. Preece and R. S. Mallozzi and C. A. Meegan and J. M. Horack and G.
J. Fishman and D. L. Band and J. L. Matteson and R. T. Skelton and J. Hakkila and L. A. Ford and C. Kouveliotou and T. M. Koshut},
  title={The Identification of Two Different Spectral Types of Pulses in Gamma-Ray Bursts},
  journal={The Astrophysical Journal},
  volume={489},
  number={1},
  pages={175},
  url={http://stacks.iop.org/0004-637X/489/i=1/a=175},
  year={1997},
  doi={	10.1086/304763}
}

@article{ hujaetal09,
	author = {{Huja, D.} and {Mészáros, A.} and {Řípa, J.}},
	title = {A comparison of the gamma-ray bursts detected  by BATSE and Swift *},
	DOI= "10.1051/0004-6361/200809802",
	url= "https://doi.org/10.1051/0004-6361/200809802",
	journal = {A&A},
	year = 2009,
	volume = 504,
	number = 1,
	pages = "67-71",
}

@article{chattopadhyayandmaitra17,
author = {Chattopadhyay, Souradeep and Maitra, Ranjan},
title = {Gaussian-mixture-model-based cluster analysis finds five kinds of gamma-ray bursts in the BATSE catalogue},
journal = {Monthly Notices of the Royal Astronomical Society},
volume = {469},
number = {3},
pages = {3374-3389},
year = {2017},
doi = {10.1093/mnras/stx1024},
URL = { + http://dx.doi.org/10.1093/mnras/stx1024},
eprint = {/oup/backfile/content_public/journal/mnras/469/3/10.1093_mnras_stx1024/2/stx1024.pdf}
}

@book{thurstone35,
  title={The Vectors of Mind: Multiple-factor Analysis for the Isolation of Primary Traits.},
  author={Thurstone, Louis Leon},
  year={1935},
  publisher={University of Chicago Press}
}

@article{maitra13,
 author = {R. Maitra},
 journal = {Sankhy\=a: The Indian Journal of Statistics, Series B},
 number = {2},
 pages = {293-318},
 publisher = {[Springer, Indian Statistical Institute]},
 title = {On the Expectation-Maximization algorithm for Rice-Rayleigh mixtures with application to noise parameter estimation in magnitude MR datasets},
 volume = {75},
 year = {2013}
}

@book{anderson03,
  title={An Introduction to multivariate statistical analysis},
  author={T. W. Anderson},
  isbn={9780471360919},
  lccn={20234317},
  series={Wiley Series in Probability and Statistics},
  year={2003},
  publisher={Wiley}
}

@article{lawley40,
title={The estimation of factor loadings by the method of maximum likelihood}, volume={60},
number={1},
journal={Proceedings of the Royal Society of Edinburgh},
publisher={Royal Society of Edinburgh Scotland Foundation},
author={D. N. Lawley},
year={1940},
pages={64–82}
}

@article{Schwarz1978,
author = {G. E. Schwarz},
year = {1978},
month = {03},
pages = {461-464},
title = {Estimating the Dimension of a Model},
volume = {6},
journal = {The Annals of Statistics}
}

@article{wilkinson1958,
author = {J. H. Wilkinson},
year = {1958},
month = {02},
title = {The Calculation of the Eigenvectors of Codiagonal Matrices},
volume = {1},
pages = {90-96},
journal = {Computer Journal},
doi = {10.1093/comjnl/1.2.90}
}

@article{day69,
 author = {N. E. Day},
 journal = {Biometrika},
 number = {3},
 pages = {463-474},
 title = {Estimating the Components of a Mixture of Normal Distributions},
 volume = {56},
 year = {1969}
}

@article{rubinandthayer82,
author = {D. B. Rubin and D. T. Thayer},
year = {1982},
month = {02},
pages = {69-76},
title = {EM algorithms for ML factor analysis},
volume = {47},
journal = {Psychometrika}
}

@book{mardiaetal06,
  title={Multivariate analysis},
  author={K. V. Mardia and J. T. Kent and J. M. Bibby},
  year={2006},
  publisher={Elsevier},
  address = {Amsterdam}
}

@article{stanek2008,
author = {Stanek, K. and Matheson, T. and Garnavich, P. and Martini, P. and Berlind, P. and Caldwell, N. and Challis, P. and Brown, Warren and Schild, Rudolph and Krisciunas, Kevin and Calkins, Michael and Lee, Janice and Hathi, Nimish and Jansen, Rolf and Windhorst, R. and Echevarria, Lizeth and Eisenstein, D. and Pindor, B. and Olszewski, E. and Bersier, David},
year = {2008},
month = {12},
pages = {L17},
title = {Spectroscopic Discovery of the Supernova 2003dh Associated with GRB 030329},
volume = {591},
journal = {The Astrophysical Journal Letters},
doi = {10.1086/376976}
}

@article{Ghirlanda2017,
author = {Ghirlanda, Giancarlo and Nappo, F. and Ghisellini, G. and Melandri, A. and Marcarini, G. and Nava, L. and Salafia, O. and Campana, S. and Salvaterra, R.},
year = {2017},
month = {10},
pages = {},
title = {Bulk Lorentz factors of Gamma-Ray Bursts},
volume = {609},
journal = {Astronomy & Astrophysics},
doi = {10.1051/0004-6361/201731598}
}

@article{berger2013,
author = {Berger, Elisha and Fong, W. and Chornock, R.},
year = {2013},
month = {06},
pages = {},
title = {An r-Process Kilonova Associated with the Short-Hard GRB 130603B},
volume = {774},
journal = {The Astrophysical Journal Letters},
doi = {10.1088/2041-8205/774/2/L23}
}

@misc{zhuetal21,
  doi = {10.48550/ARXIV.1904.06366},
  
  url = {https://arxiv.org/abs/1904.06366},
  
  author = {Zhu, Yifan and Dai, Fan and Maitra, Ranjan},
  
  title = {Visualization of Labeled Mixed-featured Datasets},
  
  publisher = {arXiv},
  
  year = {2021},
  
  copyright = {arXiv.org perpetual, non-exclusive license}
}

@article{iokaetal16,
	doi = {10.3847/1538-4357/833/1/110},
	url = {https://doi.org/10.3847/1538-4357/833/1/110},
	year = 2016,
	month = {dec},
	publisher = {American Astronomical Society},
	volume = {833},
	number = {1},
	pages = {110},
	author = {Kunihito Ioka and Kenta Hotokezaka and Tsvi Piran},
	title = {Are Ultra-long Gamma-Ray Bursts Caused by Blue Supergiant Collapsars, Newborn Magnetars, or White Dwarf Tidal Disruption Events?},
	journal = {The Astrophysical Journal}
}

@article{sorensen92,
  title={Implicit Application of Polynomial Filters in a $k$-step Arnoldi Method},
  author={Sorensen, Danny C},
  journal={{SIAM} Journal on Matrix Analysis and Applications},
  volume={13},
  number={1},
  pages={357--385},
  year={1992},
  publisher={SIAM}
}

@article{duttaandmondal15,
author = {S. Dutta and D. Mondal},
year = {2015},
month = {09},
title = {An H-likelihood Method for Spatial Mixed Linear Model Based on Intrinsic Autoregressions},
volume = {77},
pages = {699-726},
journal = {Journal of the Royal Statistical Society: Series B (Statistical Methodology)},
}

@article{byrdetal95,
  title={A Limited Memory Algorithm for Bound Constrained Optimization},
  author={R. H. Byrd and P. Lu, J. Nocedal and C. Zhu},
  journal={SIAM Journal on Scientific Computing},
  year={1995},
  volume={16},
  pages={1190-1208}
}

@misc{daietal21,
  doi = {10.48550/ARXIV.2111.04940},
  
  url = {https://arxiv.org/abs/2111.04940},
  
  author = {Dai, Fan and Dorman, Karin S. and Dutta, Somak and Maitra, Ranjan},
  
  title = {Exploratory Factor Analysis of Data on a Sphere},

  publisher = {arXiv},
  
  year = {2021},
  
  copyright = {arXiv.org perpetual, non-exclusive license}
}

@article{daietal20,
author = {Dai, Fan and Dutta, Somak and Maitra, Ranjan},
year = {2020},
title = {A Matrix-Free Likelihood Method for Exploratory Factor Analysis of High-Dimensional Gaussian Data},
journal = {Journal of Computational and Graphical Statistics},
volume = {29},
number = 3,
pages = {675-680}
}

@article{hershberger05,
author = {Hershberger, S. L.},
year = {2005},
month = {},
pages = {636-644},
title = {Factor scores},
volume = {},
journal = {Encyclopedia of Statistics in Behavioral
Science}
}

@article{bartlett37,
author = {Bartlett, Maurice},
year = {1937},
month = {04},
pages = {97 - 104},
title = {The Statistical Concept of Mental Factors},
volume = {28},
journal = {British Journal of Psychology. General Section},
doi = {10.1111/j.2044-8295.1937.tb00863.x}
}

@article{costello2005,
author = {A. B. Costello and J. Osborne},
year = {2005},
month = {01},
pages = {1-9},
title = {Best Practices in Exploratory Factor Analysis: Four Recommendations for Getting the Most From Your Analysis},
volume = {10},
journal = {Practical Assessment, Research \& Evaluation}
}

@article{chattopadhyayandmaitra18,
    author = {Chattopadhyay, Souradeep and Maitra, Ranjan},
    title = "{Multivariate t-mixture-model-based cluster analysis of BATSE catalogue establishes importance of all observed parameters, confirms five distinct ellipsoidal sub-populations of gamma-ray bursts}",
    journal = {Monthly Notices of the Royal Astronomical Society},
    volume = {481},
    number = {3},
    pages = {3196-3209},
    year = {2018},
    month = {07},
    issn = {0035-8711},
    doi = {10.1093/mnras/sty1940},
    url = {https://doi.org/10.1093/mnras/sty1940}
}

@article{chattopadhyayetal22,
  title={Multi-layered characterisation of hot stellar systems with confidence},
  author={Souradeep Chattopadhyay and Steven D. Kawaler and Ranjan Maitra},
  journal={Publications of the Astronomical Society of Australia},
  year={2022},
  volume = {39},
  number={e029},
  pages = {1-11},
  doi = {10.1017/pasa.2022.25}
}


\appendix
\section{Theoretical details regarding the MixFAD algorithm}
\label{sec:supp-meth}
As mentioned in Section~\ref{sec:fa}, the main challenge is the computationally efficient M-step estimation of $\bLambda_k$'s and $\bPsi_k$'s, so we detail our approach here that 
jointly updates $\bLambda_k$ and $\bPsi_k$ using a profile likelihood method \citep{daietal20} given the $\bSigma_{k}^*$ obtained in \eqref{eq:est}. Specifically, $\bLambda_k$ can be profiled out from the Q-function \eqref{eqn:gmm-qfun} using
\begin{result}\label{result:meth-profile}
  For a positive-definite diagonal matrix $\bPsi_k$, let $\theta_{1,k} \geq
  \theta_{2,k} \geq \cdots \geq \theta_{q,k}$, be the $q$ largest eigenvalues
  of ${\bW_k} = \bPsi_k^{-1/2}\bSigma_{k}^*\bPsi_k^{-1/2}$. Let the columns of
  $\bV_{q,k}$ store the eigenvectors corresponding to these eigenvalues. Then, the Q-function is maximised w.r.t. $\bLambda_k$ at ${\hat\bLambda_k} = \bPsi_k^{1/2}{\bV}_{q,k}\bDelta_k,$ where $\bDelta_k$ is a $q\times q$ diagonal matrix with $j$th diagonal entry $[\max(\theta_{j,k}-1,0)]^{1/2}.$ The profile Q-function is 
\begin{equation}\label{eqn:meth-profile}
\begin{split}
\mathrm{Q}_p(\bPsi_k) = c - \frac n2 \Big\{&\log\det\bPsi_k + \Tr \bPsi_k^{-1}\bSigma_{k}^*\\
 &+ \sum_{j=1}^q(\log\theta_{j,k} - \theta_{j,k} + 1)\Big\},
\end{split}
\end{equation}
where $c$ is a constant independent of $\bPsi_k$ and $\bSigma_{k}^*$ is the given estimate in \eqref{eq:est}. Further, $\nabla \mathrm{Q}_p(\bPsi_k) = -\shalf n~\mathrm{diag}({\hat{\bLambda}_k}{\hat{\bLambda}_k}^\top + \bPsi_k - \bSigma_{k}^*).$
\end{result}

\begin{proof}
The proof of Result~\ref{result:meth-profile} follows as a corollary of deriving the profile loglikelihood function in \citet{daietal20}. First, we obtain the Q-function w.r.t. $\bLambda_k,\bPsi_k$, which is a normal loglikelihood function as follows,
\begin{equation} \label{eq:exp-llk2-eq}
\begin{split}
       Q(\bLambda_k,\bPsi_k) = c 
       &-\frac n2
       \Big\{\log\det (\Lambda_k\Lambda_k^\top+\Psi_k) \\
       &\qquad\quad+\Tr (\Lambda_k\Lambda_k^\top+\Psi_k)^{-1}\bSigma_k^*\Big\},
       \end{split}
\end{equation}
where $c$ is a constant independent of $\bLambda_k,\bPsi_k$.\\
\noindent
Based on \eqref{eq:exp-llk2-eq}, the ML estimators of $\Lambda_k$ and $\Psi_k$ are obtained by solving the score equations
\begin{equation} \label{eq:system_eq-ch3}
\begin{split}
\begin{cases}
    & \Lambda_k(\mathbf{I}_q + \Lambda_k^\top\Psi_k^{-1}\Lambda_k) = 
    \bSigma_k^*\Psi_k^{-1}\Lambda_k\\
    & \Psi_k = \mathrm{diag}(\bSigma_k^*-\Lambda_k\Lambda_k^\top)
    \end{cases}
    \end{split}
\end{equation}
From $\Lambda_k(\mathbf{I}_q + \Lambda_k^\top\Psi_k^{-1}\Lambda_k) = 
   \bSigma_k^*\Psi_k^{-1}\Lambda_k$, we have
    \begin{equation}\label{eq:profileout_Lambda-ch3}
    \begin{split}
	    \Psi_k^{-1/2}\Lambda_k(\mathbf{I}_q +  (\Psi_k^{-1/2}\Lambda_k&)^\top  \Psi_k^{-1/2}\Lambda_k) \\ & = \Psi_k^{-1/2}\bSigma_k^*\Psi_k^{-1/2}\Psi_k^{-1/2}\Lambda_k.
       \end{split}
    \end{equation}
Suppose that $\Psi_k^{-1/2}\bSigma_k^*\Psi_k^{-1/2} = \bV_k\bD_k\bV_k^\top$ and that the diagonal elements in $\bD_k$ are in decreasing order with $\theta_{1,k}\geq\theta_{2,k}\geq,\cdots,\geq\theta_{p,k}$. Let $\bD_k = \begin{bmatrix} 
\bD_{q,k} & 0 \\
0 & \bD_{m,k}
\end{bmatrix}$ with $m=p-q$ and $\bD_{q,k}$ containing the largest $q$ eigenvalues  $\theta_{1,k}\geq\theta_{2,k}\geq,\cdots,\geq\theta_{q,k}$. The corresponding $q$ eigenvectors form columns of the matrix $\bV_{q,k}$ so that $\bV_k = [\bV_{q,k},\mathbf{V}_{m,k}]$. Then, if $\bD_{q,k} > \mathbf{I}_q$, \eqref{eq:profileout_Lambda-ch3} shows that
\begin{equation}
     \Lambda_k =  \Psi_k^{1/2}\bV_{q,k}(\bD_{q,k}-\mathbf{I}_q)^{1/2}.
\end{equation}
Hence, conditional on $\Psi_k$, $\Lambda_k$ is maximised at $\hat{\Lambda}_k = \Psi_k^{1/2}\bV_{q,k}\bm{\Delta}_k$, where $\bm{\Delta}_k$ is a diagonal matrix with elements $\mathrm{max}(\theta_{j,k}-1,0)^{1/2}, j = 1,\cdots,q$.

From the construction of $\mathbf{V}_{q,k}$ and $\mathbf{V}_{m,k}$, we have $\mathbf{V}^\top_{q,k}\mathbf{V}_{q,k} = \mathbf{I}_q\text{, }
\mathbf{V}^\top_{m,k}\mathbf{V}_{m,k} = \mathbf{I}_m\text{, }
\mathbf{V}_{q,k}\mathbf{V}^\top_{q,k} + \mathbf{V}_{m,k}\mathbf{V}^\top_{m,k} = \mathbf{I}_p\text{, }\mathbf{V}^\top_{q,k}\mathbf{V}_{m,k} = \boldsymbol{0}$ and hence, $(\mathbf{V}_{q,k}\mathbf{D}_{q,k}\mathbf{V}^\top_{q,k} +  \mathbf{V}_{m,k}\mathbf{V}^\top_{m,k})(\mathbf{V}_{q,k}\mathbf{D}^{-1}_{q,k}\mathbf{V}^\top_{q,k} +  \mathbf{V}_{m,k}\mathbf{V}^\top_{m,k}) = \mathbf{I}_p$.

Let $\mathbf{A}_k = \mathbf{V}_{q,k}\bm{\Delta}_k^2\mathbf{V}^\top_{q,k}$. Then $\mathbf{A}_k\mathbf{A}_k = \mathbf{V}_{q,k}\bm{\Delta}_k^4\mathbf{V}^\top_{q,k}$ and
\begin{equation}\label{eq:matrix_results1-ch3}
    \begin{split}
       |\mathbf{A}_k+\mathbf{I}_p| 
       =  |(\mathbf{A}_k+\mathbf{I}_p)\mathbf{A}_k|/|\mathbf{A}_k| 
       = & \frac{|\mathbf{V}_{q,k}(\bm{\Delta}_k^4+\bm{\Delta}_k^2)\mathbf{V}^\top_{q,k}|}{|\mathbf{V}_{q,k}\bm{\Delta}_k^2\mathbf{V}^\top_{q,k}|} \\
       = & |\bm{\Delta}_k^2+\mathbf{I}_{q,k}| = \prod_{j=1}^{q}{\theta_{j,k}}
       \end{split}
\end{equation}
and
\begin{equation}\label{eq:matrix_results2-ch3}
      \begin{split}
      (\mathbf{A}_k+\mathbf{I}_p)^{-1} 
      = & (\mathbf{V}_{q,k}\bm{\Delta}_k^2\mathbf{V}^\top_{q,k} + \mathbf{V}_{q,k}\mathbf{V}^\top_{q,k} + \mathbf{V}_{m,k}\mathbf{V}^\top_{m,k})^{-1}\\
      = & (\mathbf{V}_{q,k}(\bm{\Delta}_k^2+\mathbf{I}_q)\mathbf{V}^\top_{q,k} +  \mathbf{V}_{m,k}\mathbf{V}^\top_{m,k})^{-1}\\
      = & (\mathbf{V}_{q,k}\mathbf{D}_{q,k}\mathbf{V}^\top_{q,k} +  \mathbf{V}_{m,k}\mathbf{V}^\top_{m,k})^{-1}\\
      = & \mathbf{V}_{q,k}\mathbf{D}^{-1}_{q,k}\mathbf{V}^\top_{q,k} +  \mathbf{V}_{m,k}\mathbf{V}^\top_{m,k}.
    \end{split}
\end{equation}
Based on
\eqref{eq:exp-llk2-eq}, \eqref{eq:matrix_results1-ch3} and \eqref{eq:matrix_results2-ch3}, the profile loglikelihood is,
\begin{equation}\label{eq:proof_proposition1-ch3}
    \begin{split}
        Q_{p}(\Psi_k) 
        = c & - \frac{n}{2}\log{|\hat{\Lambda}_k\hat{\Lambda}_k^\top + \Psi_k|} \\
        &-\frac{n}{2}\Tr(\hat{\Lambda}_k\hat{\Lambda}_k^\top + \Psi_k)^{-1}\bSigma_k^*\\
        = c & - \frac{n}{2}\Big\{\log|\Psi_k^{1/2}(\mathbf{V}_{q,k}\bm{\Delta}_k^2\mathbf{V}^\top_{q,k} + \mathbf{I}_p)\Psi_k^{1/2}| \\
        &\qquad\quad +\Tr(\Psi_k^{1/2}(\mathbf{V}_{q,k}\bm{\Delta}_k^2\mathbf{V}^\top_{q,k} \\
        &\qquad\quad\qquad\quad + \mathbf{I}_p)\Psi_k^{1/2})^{-1}\bSigma_k^*\Big\}\\
        = c & - \frac{n}{2}\Big\{ \log\det\Psi_k + \log{|\mathbf{V}_{q,k}\bm{\Delta}_k^2\mathbf{V}^\top_{q,k} + \mathbf{I}_p|}  \\
        &\qquad\quad +\Tr(\mathbf{V}_{q,k}\mathbf{D}^{-1}_{q,k}\mathbf{V}^\top_{q,k} \\
        &\qquad\quad\qquad\quad +  \mathbf{V}_{m,k}\mathbf{V}^\top_{m,k})\Psi_k^{-1/2}\bSigma_k^*\Psi_k^{-1/2}
        \Big\}\\
        = c & - \frac{n}{2}\Big\{  \log\det\Psi_k + \sum_{j=1}^{q}{\log{\theta_{j,k}}} \\ &\qquad\quad +\Tr\mathbf{D}^{-1}_{q,k}\mathbf{V}^\top_{q,k}\mathbf{V}_k\mathbf{D}_k\mathbf{V}_k^\top\mathbf{V}_{q,k} \\
        &\qquad\quad + \Tr\mathbf{V}^\top_{m,k}\mathbf{V}_k\mathbf{D}_k\mathbf{V}_k^\top\mathbf{V}_{m,k}\Big\}\\
	= c & - \frac{n}{2}\Big\{  \log\det\Psi_k + \sum_{j=1}^{q}{\log{\theta_{j,k}}} + \Tr\mathbf{D}^{-1}_{q,k}\mathbf{D}_{q,k} \\
 &\qquad\quad + \Tr\mathbf{D}_{m,k}\Big\}\\
        = c &- \frac{n}{2}\Big\{ \log\det\Psi_k + \Tr\Psi_k^{-1}\bSigma_k^* + \sum_{j=1}^{q}{\log{\theta_{j,k}}} \\
        &\qquad\quad -\sum_{j=1}^{q}{\theta_{j,k}} + q \Big\},
    \end{split}
\end{equation}
where $c$ is a constant free of $\Psi_k$.
\end{proof}
From Result~\ref{result:meth-profile}, we need the $q$ largest eigenvectors of $\bW_k$. 
The partial eigendecomposition results of $\bW_k$ can be efficiently obtained via the Lanczos algorithm \citep{sorensen92,duttaandmondal15} that involves $\bW_k$ through only the matrix-vector products and hence sufficiently reduce the storage and computational time. (See the detailed Lanczos algorithm in the next section.) Then, we can obtain $\bPsi_{k}^* = \arg\max \mathrm{Q}_p(\bPsi_k)$ via the limited-memory Broyden-Fletcher-Goldfarb-Shanno
quasi-Newton algorithm \citep{byrdetal95} with box-constraints (L-BFGS-B). This algorithm uses values of
$Q_p(\Psi_k)$ and $\nabla Q_p(\Psi_k)$ from the last few iterations to approximate the exact Hessian matrix,
reducing storage costs from $O(p^2)$ to $O(p)$. Given $\bPsi_{k}^*$, the loading matrix is updated as $\bLambda_{k}^* = \bPsi_{k}^{*1/2}{\bV}_{q,k}\bDelta_k$.

\subsection*{Lanczos algorithm for the partial eigendecomposition of $\bW_k$}
As illustrated by \citet{daietal21}, the $q$ largest eigenvalues and eigenvectors of ${\bW}_k = \bPsi_k^{-1/2}\bSigma_{k}^*\bPsi_k^{-1/2}$ can be obtained using the implicitly restarted Lanczos algorithm \citep{sorensen92}. Suppose that $m = \max\{2q + 1, 20\}$ and that $\mathbf{f}_{1,k} \in \bbR^p$ is any vector with $\|\mathbf{f}_{1,k}\| = 1$ and initialize $\mathbf{F}_{1,k} = \mathbf{f}_{1,k}.$
We then employ the Lanczos iterations \citep{duttaandmondal15} as follows.
For $l=1,2,\ldots,m$,
\begin{itemize}
\item Compute ${\bu}_{l,k} = {\bW_k\bff}_{l,k}$ and $\alpha_{l,k} = {\bff}_{l,k}^\top{\bu}_{l,k}$.
\item Compute ${\br}_{l,k} = {\bu}_{l,k} - \alpha_{l,k}{\bff}_{l,k} - \beta_{l-1,k}{\bff}_{l-1,k}$ (assuming $\beta_{0,k}=0$ and ${\bff}_{0,k} = {\bzero}$).
\item Let $\beta_{l,k} = \|{\br}_{l,k}\|$ and if $l<q$ and $\beta_{l,k}\ne 0,$ compute ${\bff}_{l+1,k} = {\br}_{l,k}/\beta_{l,k}$ and set ${\bF}_{l+1,k}=[{\bF}_{l,k},{\bff}_{l+1,k}]$.
\end{itemize}
Suppose that ${\bT}_{m,k}$ is the $m\times m$ symmetric tridiagonal
matrix with diagonal entries $\alpha_{1,k},\alpha_{2,k},\ldots,\alpha_{m,k}$ and $j$th off-diagonal 
entries $\beta_{j,k}$ for $j=1,\ldots,m-1$.
We compute the eigenvalues $e_{1,k} > e_{2,k} > \cdots > e_{m,k}$ of ${\bT}_{m,k}$ with eigenvectors ${\bg}_{1,k},{\bg}_{1,k},\ldots,{\bg}_{m,k}$ via a Sturm sequencing algorithm \citep{wilkinson1958}. Also let ${\bv}_{j,k} = {\bF}_{m,k}{\bg}_{j,k},$ for $1\leq j \leq m.$ The $e_{j,k}$'s and ${\bv}_{j,k}$'s are called Ritz values and Ritz vectors of ${\bW}_k$. It can be shown that $\|{\bW_k}{\bv}_{j,k} - {\bv}_{j,k}e_{j,k}\| = \beta_{m,k}|g_{m,j,k}|$, where $g_{m,j,k}$ is the $m$th entry of vector ${\bg}_{j,k}$, for $j=1,2,\ldots,m.$ The algorithm stops if
\begin{equation}\label{eqn:lanczosStop}
 \beta_{m,k}\max_{1\leq j \leq m}|g_{m,j,k}| < \delta
\end{equation}
for some prespecified tolerance $\delta$ and $e_{1,k},e_{2,k},\ldots,e_{q,k}$ and ${\bv}_{1,k},\bv_{2,k},\ldots,{\bv}_{q,k}$ are accurate approximations of the $q$ largest eigenvalues and corresponding eigenvectors of ${\bW}_k$.

However, in practice, more iterations are needed for the Ritz vectors
and Ritz values to converge to the eigenvalues and eigenvectors of ${\bW}_k$, so~\citet{sorensen92} suggests implicitly restarting the
Lanczos algorithm and also shifting the spectrum of the symmetric tridiagonal matrices iteratively to force
the new residuals ${\br}_{m,k}$ to zero, thereby accelerating the
convergence rate. So we 
compute the QR-decompositions: ${\bT}_{m,k} - e_{j,k}{\bI}_m = {\bQ}_{j,k}{\tilde{\bR}}_{j,k},$ for $j=q+1,\ldots,m,$ let $\tilde{{\bQ}}_k = {\bQ}_{q+1,k}{\bQ}_{q+2,k}\cdots{\bQ}_{m,k}$ and reset
${\bF}_{m,k} = {\bF}_{m,k}\tilde{{\bQ}}_k$ and ${\bT}_{m,k} = \tilde{{\bQ}}_k^\top
{\bT}_{m,k}\tilde{{\bQ}}_k.$ Then 
\begin{equation}\label{eqn:restart}
 {\bW\bF}_{q,k} = {\bF}_{q,k}{\bT}_{q,k} + \beta_k^*{\bff}_{q+1,k}{\be}_{q}^\top
\end{equation}
where $\beta_k^*$ is the $(q+1,q)$th entry of ${\bT}_{m,k},$ ${\be}_{q}$ is
the $q$th canonical basis vector in $\bbR^{q},$ and ${\bT}_{q,k}$ is the
$q\times q$ principal sub-matrix of ${\bT}_{m,k}$~\citep{sorensen92}. Therefore,
\eqref{eqn:restart} is itself a $q$th-order Lanczos factorization of
${\bW}_k.$ Next, we ``restart'' the Lanczos iterations from $l=q+1,\ldots,m$ instead of $1$ through $m$, terminating if \eqref{eqn:lanczosStop} is satisfied, and restarting the algorithm otherwise. 

\bsp	
\label{lastpage}
\end{document}